\documentclass[11pt,a4paper]{article}
\pdfoutput=1

\usepackage{jheppub}

\usepackage{amsmath}
\usepackage{amssymb}
\usepackage{mathtools}
\usepackage{physics}
\usepackage{slashed}
\usepackage{simpler-wick}

\usepackage{microtype}
\usepackage{dsfont}
\usepackage{eufrak}
\usepackage{bbold}

\usepackage{graphicx}
\usepackage{float}
\usepackage{subcaption}
\usepackage[export]{adjustbox}
\usepackage{wrapfig}

\usepackage{tikz}
\usepackage[tikz]{mdframed}

\mdfdefinestyle{proofstyle}{outerlinewidth = 0.2 , roundcorner = 2pt , leftmargin = 4 , rightmargin = 4 , backgroundcolor = gray!15 , linecolor = gray!25 , innertopmargin = 10pt , innerbottommargin = 10pt , everyline = true , splittopskip = 18pt , splitbottomskip = 10pt}

\makeatletter
\renewcommand*\env@matrix[1][\arraystretch]{%
  \edef\arraystretch{#1}%
  \hskip -\arraycolsep
  \let\@ifnextchar\new@ifnextchar
  \array{*\c@MaxMatrixCols c}}
\makeatother

\title{
Spinors in (Anti-)de Sitter Space 
}
\author{Vladimir Schaub}
\emailAdd{vladimir.schaub@kcl.ac.uk}
\affiliation{Mathematics Department, King's College London, \\
The Strand, London,  WC2R 2LS, UK}
\abstract{
 We explore analytical aspects of correlators involving Dirac spinors in $d+1$-dimensional de Sitter space. Adapting the formalism of Sleight and Taronna, we show how to relate processes involving fermions in the in-in formalism to equivalent Witten diagrams in (complexified) Euclidean anti-de Sitter space. We exemplify the method for a fermion-exchange diagram. We establish a positive spectral decomposition over the principal series of the Wightman function of two spinors, showing the consequences of unitarity.
}

\makeatletter
\def\@fpheader{\vspace{0cm}}
\makeatother

\begin{document}
\maketitle

\section{Introduction}

The experimental facts motivating this theoretical work are the observations that neutrinos exist and space expands. It is of crucial importance then to have a well developed understanding of the physics of relativistic field of spin $1/2$ in spacetimes which are asymptotically de Sitter (dS). The goal of the present paper is to contribute to this edifice, by answering the question : how does one computes cosmological correlators involving Dirac fermions in pure dS$_{d+1}$ ? 

This endeavour is an incremental piece to a vast existing amount of literature. Inflationary cosmologists have studied for a long time the properties of cosmological correlators, which are the expectation values of field insertions at the late time surface of de Sitter, as probes of the early universe\cite{Arkani-Hamed:2015bza}, for a review see \cite{Baumann:2009ds,Green:2022hhj}. Because of the difficulty in analytically evaluating these processes, studies has turned toward more axiomatic approaches. Some focused on bootstrapping the correlators, in works such as \cite{Arkani-Hamed:2018kmz,Baumann2018,Baumann:2020ksv,Baumann:2020dch,Baumann:2019oyu}, reviewed generally in \cite{Baumann:2022jpr}. Others have followed an holographic perspective \cite{McFadden:2009fg,McFadden:2010na,McFadden:2010vh,McFadden:2011kk,Bzowski:2012ih}. These are of course not the sole interesting objects in expanding spaces. Many studies have investigated the rich physics of quantum fields living in de Sitter, using an array of tools beyond late time correlators only, such as \cite{Bros:1994dn,Strominger:2001pn,Anninos:2011ui,Gorbenko:2019rza,Hogervorst:2021uvp}. For a more thorough review, see \cite{Flauger:2022hie}. 

The present work is motivated through the recent renewed activity on the analytical front, started through the work of Sleight and Taronna \cite{Sleight:2019hfp,Sleight:2019mgd,Sleight:2020obc,Sleight:2021plv}, which have offered a new analytical approach to computing correlators. The crucial insight they offered is that one is able to rewrite Lorentzian processes in dS as equivalent processes in Euclidean Anti-de Sitter (AdS) space, which then allows for an evaluation using standard methods. The Wick rotation between these spaces \cite{Maldacena:2002vr} is there explicitly realised. This has led to various interesting exact computations and results, and more generally to the import of AdS knowledge into the properties of perturbative dS systems \cite{Fichet:2021xfn,DiPietro:2021sjt,Heckelbacher:2022hbq,Sleight:2021iix}.

The goal of this article is to close the current gap in these development relative to spinorial insertions. Our first main result, presented in sec. \ref{sec:ess}, is a new set of rules to rewrite cosmological diagrams involving fermions using an equivalent Witten diagram, expanding the ones of \cite{Sleight:2021plv}. Regardless of any inflationary applications, from a physical and formal standpoint these are perfectly legitimate objects to understand. This work is a continuation of our previous study on fermions in de Sitter space \cite{Pethybridge:2021rwf}, which investigated the embedding formalism and showed in detail how to uplift and project spinors, thereby streamlining their manipulations and giving a simple and efficient derivation of their Wightman functions. Work on spinors in expanding spaces goes back to Dirac \cite{Dirac:1935zz}, who attempted to formulate them in the embedding space already. Other studies such as \cite{Candelas1975,Fang:1979hq,Allen:1986qj,Cotaescu2002,Cotaescu2018} have considered them, and many more focused on the relation to field theory defined on a sphere, such as \cite{Camporesi1992,Camporesi1996,Letsios:2020twa}. Our work is framed as a bridge between these considerations and the computation of cosmological correlators involving Dirac particles. As this endeavour makes us consider in detail the analytical structure of fermion correlators, it is natural to wonder as well about the more general properties of these objects, such as unitarity \cite{Bros:1990cu,Bros:1994dn,Bros:1995js,Bros:1998ik,Bros:2010rku,Sengor:2019mbz,DiPietro:2021sjt,Hogervorst:2021uvp}. To this aim, we also prove that spinors admit a positive Källen-Lehman decomposition in de Sitter space. This forms our second main result, presented in sec. \ref{sec:kl}

The plan of the paper is as follows. We first review the existing results for the analytical continuation of bosonic cosmological correlators. Our approach focuses on the differential relations defining the different propagators, while working in the embedding picture. The analytical continuation rules are specified in sec. \ref{sec:anat}. The second section discusses the analogue for fermions. As fermions are fickle creatures very sensitive to signature, we setup in detail the conventions in both the embedding and real spaces. A careful analytical continuation follows, from which Feynman rules are extracted. As a proof of concept, we compute a sample cosmological diagram in sec. \ref{sec:comput} . The final section is independent from the rest of the paper, and focuses on the spectral decomposition. We review the canonical quantisation of scalar and Dirac fields in dS, resumming the propagator from the mode functions. We consider generic on-shell states, and show that symmetries fix the overlap of the Dirac field with them to be proportional to the free-field modes. A positive decomposition then ensues. Our conventions follow the first appendix of \cite{Pethybridge:2021rwf}. Our appendix is dedicated to a detailed rederivation of the properties of spinorial harmonic functions.

\section{Euclidean AdS Method for Bosonic Cosmological Correlator}

In this section, we are concerned with how to compute cosmological correlation functions by relating them to Witten diagrams. The content is not new, forming an alternative exposition of some of the existing literature \cite{Sleight:2020obc}. However, unlike some of these works, we do not use Mellin space. Our goal is to set the conventions and procedure in a way that makes the transition to the spinor case smooth. The observables of interest are $n$-point functions of field inserted on a fixed late-time slice, $\eta_c \approx 0^{-}$. Because of the $SO(1,d+1)$ symmetry present at the late-time boundary, this must take the form of a CFT $n$-point function. In principle, for a given Lagrangian QFT in dS, this correlator can be derived from perturbation theory. The relevant scheme is the Schwinger-Keldysh, or in-in formalism, which uses different type of propagators \cite{Weinberg2005,Adshead2009}.

We first review the embedding approach to study field insertions in (EA)dS  \cite{Costa:2014kfa,Pethybridge:2021rwf}, and its result for the general Wightman function of symmetric traceless tensor (STT) operators in (EA)dS. We then review the in-in formalism, and its realisation using $i\epsilon$ prescription. We discuss how to wick rotate field insertions from dS to EAdS \cite{Sleight:2019hfp,Sleight:2020obc,DiPietro:2021sjt}, and check how the analytically continued propagators decompose in term of AdS propagators. Finally, we sum up the resulting procedure in a series of Feynman rules, like in \cite{Sleight:2021plv}. 

\subsection{Fields in the Embedding Space} 

\begin{wrapfigure}{r}{0.5\textwidth}
	\centering
	\includegraphics[width=0.8\linewidth]{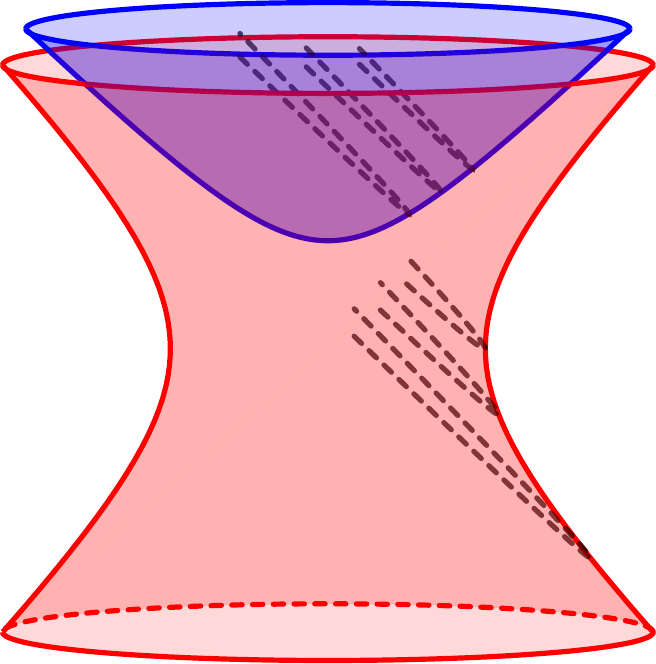}
	\caption{The embedding of dS$_{d+1}$ (red) and EAdS$_{d+1}$ (blue) into $\mathbb{R}^{1,d+1}$. Constant $\eta$ and $z$ coordinate lines are drawn in black. These get deformed smoothly one into another through analytical continuation, bypassing the light-cone.}
	\label{fig:embed}
	\vspace*{-0.8cm}
\end{wrapfigure}
The building blocks of perturbation theory in (EA)dS are the Wightman functions of elementary fields. These objects are determined by the isometry group and physical requirements. To take advantage of this fact, it is convenient to consider them in a formalism that fully exploit the available symmetries, the embedding picture.

Consider dS$_{d+1}$ as the subset $X^{A} \in \mathbb{R}^{1,d+1}$ satisfying $X^{A}X_A = +1$, where we measure in unit of the radius of de Sitter. A coordinate system which covers part of the hyperboloid is given through the Poincaré coordinates 
\begin{equation}
\begin{aligned}\label{eq:poincare}
	X^{A} &= (X^{+},X^{-},X^{a}) \\
	&=\frac{1}{\eta}(1,x^2-\eta^2,x^{a}) \, ,
\end{aligned}
\end{equation}
with $\eta<0$, and $X^{\pm}=X^{0}\pm X^{d+1}$. One can study EAdS$_{d+1}$ insertions in this embedding space as well, by looking at points satisfying $X_{H}\cdot X_H = -1$, which can be parametrised through
\begin{align}
	X_H^{A} = \frac{1}{z}(1,x^2+z^2,x^{a}) \, ,
\end{align}
with $z>0$. The situation is represented in fig.\ref{fig:embed}, with coordinate lines drawn.

In this augmented space, the isometries are restated in term of Lorentz-invariance, which is simple to impose. One can define the uplift and projection of tensorial fields from $dS_{d+1}$ to and from the embedding space $\mathbb{R}^{1,d+1}$, using the usual rules of differential geometry \cite{bruhat:1982}. As before, the analogous statement hold for EAdS. A central element of this analysis is the covariant derivative in dS, which in the embedding becomes
\begin{align}
	\nabla_A = \partial_A - X_{A} X\cdot \partial \, ,
\end{align}
And similarly for EAdS by switching the sign. The canonical example to showcase the mileage brought by this formalism is that of the free massive scalar. Consider the Wightman function of scalar fields in dS$_{d+1}$, which obey the equations of motions
\begin{align}
	(\nabla_A \nabla^{A}-m^2)\expval{\phi(X)\phi(Y)}=0 \, .
\end{align}
From Lorentzian symmetry, this is only a function of the scalar cross-ratio 
\begin{align}\label{eq:crossratio}
	u=\frac{1+X\cdot Y}{2} =\frac{(\eta_1+\eta_2)^2-(x-y)^2}{4\eta_1\eta_2} \, .
\end{align}
For points in AdS, the cross-ratio is defined identically in the embedding. We can apply the equation of motions on this ansatz and solve the resulting differential equation,
\begin{align}\label{eq:w0}
	\expval{\phi(X)\phi(Y)}&= W_{0,\Delta}(u)= \frac{\Gamma(\Delta)\Gamma(d-\Delta)}{(4\pi)^{\frac{d+1}{2}}\Gamma(\frac{d+1}{2})} \,_2 F_1 \left(d-\Delta,\Delta;\frac{d+1}{2};\frac{1+X\cdot Y}{2} \right)  \, ,
\end{align}
where we used the previously mentioned physical requirements, namely the Bunch-Davies and Hadamard conditions. The Bunch-Davies condition selects a vacuum state that maps to the Minkowski one in the far past limit, while the Hadamard condition relates the normalisation of this correlator to the one in flat space using the coincident point divergence. One can in principle consider more general $\alpha$-vacua as well \cite{Allen:1985ux,Sasaki:1994yt,Bousso:2001mw}, but for simplicity we will not consider that complication in the present work. The range of $\Delta$ allowed by unitarity is an interesting matter, which depends non-trivially on the spin of the particle, and is made of different series. The details are beyond the scope of our work, and we refer the reader to the abundant literature touching on the subject, such as \cite{Joung2006,Joung2007,Anous2020,Sun2021,Penedones:2023uqc}. 

This Wightman function can be used to construct the different propagators which are needed to compute diagrams in the Schwinger-Keldysh framework.

The previous computation can be generalised to the case of tensorial operators \cite{Costa:2011wa,Costa:2014kfa}. Consider the index-free field $F_{J,\Delta}(X,W)=W^{A}\ldots W^{A_J}F_{A_1 \ldots A_J}(X)$. This homogeneous polynomial in the polarisation variables $W^{A}$ encodes a spin-$J$ symmetric tensor in the embedding. Requiring $W^{A}W_{A}=0$ selects out traceless tensors. Further requiring $W\cdot X=0$, we restrict ourselves to transverse tensors in the embedding, which under the pull-back map to the dS slice, these are in one-to-one map with STT fields. As in the scalar case, the ambient symmetry fixes the ansatz for the Wightman function, on which one can solve the equations of motions
\begin{equation}
\begin{aligned}
	\expval{F_{J,\Delta}(X_1,W_1)F_{J,\Delta}(X_2,W_2)}&=W_{J,\Delta}\left(u\right) \\
	&=\sum_{k=0}^{J}(W_1\cdot W_2)^{J-k}(W_1\cdot X_2 W_2 \cdot X_1)^{k}g_{k}(u) \, .
\end{aligned}
\end{equation}

The Wightman function $W_{J,\Delta}$ is determined by the functions $g_{k}(u)$. The situation is the same in EAdS, and we call the analogous propagator $\Pi_{J,\Delta}(u)$  \cite{Costa:2014kfa}. The functions $g_k$ are specified through the relation 
\begin{equation}
	g_{k}(u)=\sum_{l=k}^{J}\left(\frac{l!}{k!}\right)^2\frac{\partial_u^{k}h_{k}(u)}{2^{l+k}(l-k)!} \, .
\end{equation} 
with the $h_k$ recursively computed through
\begin{equation}
\begin{aligned}\label{eq:hk}
h_{k}(u)&= \begin{aligned}[t]c_{J,\Delta,k} \bigg( &2(d-2k+2J-1)\bigg((d+J-2)h_{k-1}(u) \\
&-\left(\frac{1}{2}-u\right)\partial_u h_{k-1}(u)\bigg)-4(2-k+J)h_{k-2}(u) \bigg) \, , \end{aligned}  \\
c_{J,\Delta,k}&=\frac{J-k+1}{k (d+2 J-k-2) (\Delta +J-k-1) (d-\Delta +J-k-1)} \, .
\end{aligned}
\end{equation}

Given a seed $h_{0}(u)$, these relations fix completely the 2-point function. Surprisingly, they hold identically in both EAdS and dS, with only the seed differing. In both cases, it is given by the Wightman function of scalars. For fields in dS, we have
\begin{align}
	h^{dS}_{0}(u)&=\frac{\Gamma(\Delta)\Gamma(d-\Delta)}{(4\pi)^{\frac{d+1}{2}}\Gamma(\frac{d+1}{2})}\,_2 F_1\left(\Delta,d-\Delta;\frac{d+1}{2};u\right) \, ,
\end{align}
while for EAdS fields, it is the analogous result, 
\begin{align}
	h^{H}_{0}(u)&=\frac{\Gamma(\Delta)}{2\pi^{d/2}\Gamma(\Delta+1-\frac{d}{2})}\frac{1}{(-4u)^{\Delta}} \,_2 F_1\left(\Delta,\Delta-\frac{d-1}{2};2\Delta-d+1;\frac{1}{u}\right) \, .
\end{align}
These functions are related in that, as we are gonna see, the first one can be written as a linear combination of the second, with conformal weight $\Delta$ and $d-\Delta$. They are found by solving the Casimir equation on the appropriate spaces and imposing singular behaviour.

With these kinematical objects fixed, one is in principle able to write diagrams in both spaces, in a given perturbative framework.

\subsection{In-In Formalism and Analytical Continuation to AdS}\label{sec:anat}

The rules to compute correlators in de Sitter are rather intricate. The objects of interest are correlation functions of fields inserted on a given time-slice $\eta_c \sim 0^{-}$. Having settled on a set of interaction vertices, we write all possible diagrams by inserting vertices decorated with $l$ and $r$ labels. The propagators to be used depend on the type of insertions, bulk or boundary, and on the label of the vertices it is joining in the bulk. The different labels indicate the relative time ordering of the operators, which is implemented through an $i\epsilon$ ($\epsilon>0$) prescription \cite{Weinberg2005}. The prescription can be phrased in term of the cross-ratio on which the 2-point function depends \cite{DiPietro:2021sjt}
\begin{align}
\begin{aligned}
	(l,l) &\Rightarrow u+i \epsilon \, , \\
	(r,r) &\Rightarrow u-i \epsilon \, ,
\end{aligned} && 
\begin{aligned}
	(l,r) &\Rightarrow u+i \epsilon \ \text{sgn}(\eta_1-\eta_2)  \, , \\
	(r,l) &\Rightarrow u-i \epsilon \  \text{sgn}(\eta_1-\eta_2) \, .
\end{aligned}
\end{align} 

The different propagators $W^{ab}(u)$ are given by evaluating the Wightman function we previously computed, at the respective arguments $u_{ab}$. This differs from EAdS. The theory is Euclidean and one works directly in perturbation theory using the Wightman function. The integrals over vertices encountered there are also under better analytical control.

This situation is not unlike usual quantum field theory, where the Lorentzian diagrams are better understood as the analytical continuation of Euclidean diagrams, which are better behaved. The procedure we now explain is the analogue Wick rotation for de Sitter space, whereby one recasts cosmological correlators in term of perturbative Euclidean AdS process \cite{Sleight:2019hfp}. This procedure hinges on a contour deformation, which must be done on a continuous path that does not cross any branch cut or pole.

\begin{wrapfigure}{r}{0.5\textwidth}
  \centering 
     \includegraphics[width=0.9\linewidth]{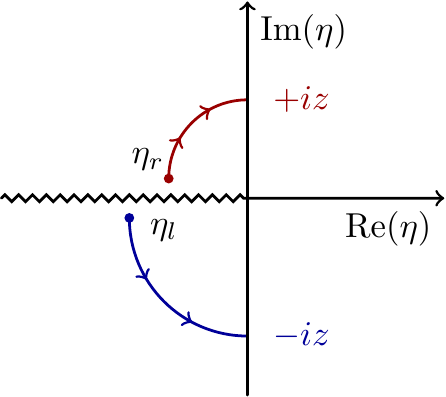}
  \captionof{figure}{Analytical Continuation. The branch-cut should not be crossed as it separates the different time orderings.}
  \label{fig:cont}
  \vspace*{-0.2cm}
\end{wrapfigure}

Consider a generic term arising from perturbation theory. The expressions are all analytic in the variable $\eta_{l/r}$ in the domain which preserve the ordering of the operator. Consider the path that rotates the time coordinate depending on its $l/r$ label along  
  \begin{equation}\begin{aligned}\label{eq:cont}
	 \eta_r &\rightarrow e^{-i\pi/2}z \, , \\ 
	 \eta_l &\rightarrow e^{+i\pi/2}z \, .
	\end{aligned}\end{equation}
From the $i\epsilon$ prescriptions spelled out before, these do not cross any singularity, hence are valid contour deformations of the different integrals one encounters \cite{Visser:2017atf}. The situation is illustrated in fig.\ref{fig:cont}. Through this transformation, the propagators and integration over coordinates are analytically continued. From the Poincaré parametrisation of the EAdS and dS slices, these translate also into an analytical continuation of the embedding coordinates. For instance, consider the integral of a point in the left contour, over all of dS,

\begin{equation}
\begin{aligned}
	i\int_{dS} D X_l = i\int d^{d+2}X \,  2\delta(X^2-1) &\rightarrow e^{i\frac{\pi}{2}(d+3)}\int d^{d+2} X  \, 2\delta(X^2+1)\\
	&=e^{i\frac{\pi}{2}(d-1)}\int_{H} D X \, ,
\end{aligned}
\end{equation}
Where we notice the last integral becomes an integral over the EAdS slice. One might complain that this is an integral over both the future and past EAdS slices. However, the integrals which appear in perturbative computation of cosmological diagrams are not integrals over the full de Sitter space, but only the causal past of the late time slice, which are the coordinate covered by the Poincaré parametrisation \eqref{eq:poincare}. As such, it is sensible to invert the logic and think of the integral over a single EAdS hyperboloid as defining the relevant integral over dS. The analogous computation for $X_r$ gives the expected result 
\begin{align}
	-i\int_{dS} D X_r& \rightarrow e^{-i\frac{\pi}{2}(d-1)}\int_{H} D X \, ,
\end{align}
and allows us to rewrite generic vertices integrals in dS as vertices in EAdS. The job is done, as we have yet to consider the behaviour of the propagators, which is the topic of the next section.

\subsection{Prolonging the In-In Propagators}

The goal of this section is to demonstrate that a given STT dS field can be rewritten as a (shadow)-pair of AdS fields of the same type, up to phase factors, effectively transcribing perturbation theory from one space to another. The starting point is the previously highlighted fact that the analytic continuation acts on the embedding coordinate by sending 
\begin{align}\label{eq:veccont}
	X_a \cdot Y_b &\rightarrow e^{-i\frac{\pi}{2}(a+b)}X \cdot Y \, ,
\end{align}
under which $u$ behaves as 
\begin{align}
	u_{ll}& \rightarrow 1-(u-i\epsilon) \, , & u_{rr}& \rightarrow 1-(u+i\epsilon) \, , & u_{lr}&\rightarrow u \, , &  u_{rl}&\rightarrow u \, .
\end{align}
In the $ll$ and $rr$ scenarii, the phases associated to terms such as $u^{\Delta}$, since $u<0$ for points in AdS, are resolved by considering in which direction in the complex plane one approaches, as we illustrated by keeping the factors of $\epsilon$. The propagators of tensors also contain structures, which are built using product of polarisation and position vectors. These transform like \eqref{eq:veccont} as well.

We now take the propagator along the paths \eqref{eq:cont} in the complex plane. Going one step further, these analytically continued dS propagators are decomposed into EAdS propagators. The matter is essentially solved by the application of the hypergeometric identities \cite{gradshtein}
\begin{align}
	\,_2 F_1\left(a,b;c;z\right)&=\begin{aligned}[t]\label{eq:id1}&\frac{\Gamma(c)\Gamma(b-a)}{\Gamma(b)\Gamma(c-a)}\frac{1}{(-z)^{a}}\,_2 F_1\left(a,a+1-c;a+1-b;\frac{1}{z}\right)\\
	&+\frac{\Gamma(c)\Gamma(a-b)}{\Gamma(a)\Gamma(c-b)}\frac{1}{(-z)^{b}}\,_2 F_1\left(b,b+1-c;b+1-a;\frac{1}{z}\right) \, , \end{aligned} \\
	\,_2 F_1\left(a,b;c;z\right)&=\begin{aligned}[t]\label{eq:id2}&\frac{\Gamma(c)\Gamma(b-a)}{\Gamma(b)\Gamma(c-a)}\frac{1}{(1-z)^{a}}\,_2 F_1\left(a,c-b;a+1-b;\frac{1}{1-z}\right)\\
	&+\frac{\Gamma(c)\Gamma(b-a)}{\Gamma(b)\Gamma(c-a)}\frac{1}{(1-z)^{b}}\,_2 F_1\left(b,c-a;b+1-a;\frac{1}{1-z}\right) \, ,
\end{aligned}
\end{align}
to the seed of the spinning propagators, as we now illustrate. Consider $W_{J,\Delta}^{lr}(u)$; it is unchanged when performing the analytic continuation, and one can reduce by linearity its decomposition to that of the seed, which is readily done using \eqref{eq:id1}. This only works thanks to the relations \eqref{eq:hk} being unchanged for $\Delta$ and $d-\Delta$. The analytic continuation of $h_{0}$ implies the analytic continuation of the whole propagator,
\begin{align}
	W_{J,\Delta}^{lr}(u) \rightarrow \frac{1}{2\sin(\frac{\pi(d-2\Delta)}{2})}\bigg(\Pi_{J,\Delta}(u)-\Pi_{J,d-\Delta}(u) \bigg) \, .
\end{align}
Which is a way of saying that $W^{lr}$ is proportional to the AdS harmonic function. $W^{rl}$ is identical. The cases $ll$ and $rr$ are more interesting. Consider the scalar propagator, or seed, using the second identity \eqref{eq:id2}, 
\begin{equation}
\begin{aligned}
	W_{0,\Delta}^{ll}(u)&\rightarrow W_{0,\Delta}(1-u) \\
	&=\frac{1}{2\sin(\frac{\pi(d-2\Delta)}{2})}\bigg(\Pi_{0,\Delta}(u)e^{i\pi \Delta}-\Pi_{0,d-\Delta}(u)e^{i\pi(d-\Delta)}\bigg) \, ,
\end{aligned}
\end{equation}
and likewise for $W_{0,\Delta}^{rr}$, with opposite phases. Having seen how the relation holds for $J=0$, we can work our way through \eqref{eq:hk}, noticing that the analytical continuation leaves the relations relating the different $h_k$ invariant, and allows us to extrapolate the decomposition of $h_0$ into the one of $W_{J}$. Explicitly,  
\begin{align}
	W_{J,\Delta}^{ll}(u) \rightarrow \sum_{k=0}^{J}(W_1\cdot W_2)^{J-k}(W_1\cdot X_2 W_2 \cdot X_1)^{k}\underbrace{(-1)^{J-k}g_{k}(1-u)}_{\bar{g}_k(u)} \, ,
\end{align}
and since $\partial_{(1-u)}f(u)=-\partial_u f(u)$, we find
\begin{align}
	\bar{g}_{k}(u)=\sum_{l=k}^{J}\left(\frac{l!}{k!}\right)^2\frac{1}{2^{l+k}(l-k)!}\partial_u^{k}\underbrace{\left(h_{k}(1-u)(-1)^{J}\right)}_{\bar{h}_{k}(u)} \, .
\end{align}
The crucial property is now that the recursive relation \eqref{eq:hk} for $h_{k}(u)$ is invariant under the change to $\bar{h}_{k}(u)=(-1)^{k}h_{k}(1-u)$, as well as under $\Delta\rightarrow d-\Delta$. It then follows that the analytical continuation of $W_{0,\Delta}^{ll}$ is indeed sufficient to find the relation for the general spin, as claimed,
\begin{align}
	W_{J,\Delta}^{ll}(u) \rightarrow \frac{1}{2\sin(\frac{\pi(d-2\Delta)}{2})}\bigg(\Pi_{J,\Delta}(u)e^{i\pi(J+\Delta)}-\Pi_{J,d-\Delta}(u)e^{i\pi(J+d-\Delta)}\bigg) \, .
\end{align}
For the $rr$ propagator one proceeds identically, finding opposite phases.

From this result, we can rewrite all propagators of tensorial operators in the in-in formalism. Notice that the four different decompositions found for the de Sitter propagator can be restated as a change of basis to new AdS fields 
\begin{align}
	F_{J,\Delta}^{dS,l} &= e^{i\frac{\pi}{2}(J+\Delta)}F^{+}_{J}+e^{i\frac{\pi}{2}(J+d-\Delta)}F^{-}_{J} \, , & F_{J,\Delta}^{dS,r} &= e^{-i\frac{\pi}{2}(J+\Delta)}F^{+}_{J}+e^{-i\frac{\pi}{2}(J+d-\Delta)}F^{-}_{J} \, ,
\end{align}
having a non-standard kinetic term normalisation 
\begin{align}
	\expval{F_{J,\Delta}^{+}F_{J,\Delta}^{+}}&=\frac{\Pi_{J,\Delta}}{2\sin(\frac{\pi(d-2\Delta)}{2})}\, , & 	\expval{F_{J,\Delta}^{-}F_{J,\Delta}^{-}}=-\frac{\Pi_{J,d-\Delta}}{2\sin(\frac{\pi(d-2\Delta)}{2})} \, .
\end{align}

Effectively, this gives a procedure to compute a generic diagram in dS. One first draws the diagram, then assign weights $\Delta_{+}=\Delta$ or $\Delta_{-}=d-\Delta$ to each line whose weight is not fixed. For each internal line, one divides by the normalising factor $\frac{1}{2\sin(\frac{\pi}{2}(d-2\Delta_\pm))}$. For each vertex, one must fix the label $l/r$, multiply by the factors of $e^{\pm i\frac{\pi}{2}(J+\Delta)}$ for each spin $J$, dimension $\Delta$ line attached to it, multiply by $e^{\pm i\frac{\pi(d-1)}{2}}$, and finally sum. This is a straightforward set of rules to relate cosmological diagrams to Witten diagrams.

It is important to match the analytical continuation of the external points as well. In general in AdS, to take the boundary limit, one simply rescales a point and the field insertion there to infinity \cite{Harlow:2011ke}. From the operator perspective, we define 
\begin{align}
	\lim_{\lambda \to \infty} \lambda^{\Delta}\phi_{H}\left(\lambda P+\frac{1}{\lambda}\ldots\right)&=\widehat{\phi}(P)  \, .
\end{align}
The boundary field $\widehat{\phi}$ is non-canonically normalised and obey a CFT transformation law for a primary of dimension $\Delta$. We define the bulk-to boundary-propagator in AdS
\begin{align}
	\lim_{\lambda \to \infty} \lambda^{-\Delta}\Pi_{\Delta}\left(\lambda P+\frac{1}{\lambda}\ldots, Y\right) = K_{\Delta}(P,Y) \, .
\end{align} 
For the scalar field
\begin{align}
	K_{\Delta}(P,Y) &=\frac{\Gamma (\Delta )}{2\sqrt{\pi}^{d}\Gamma \left(\Delta-\frac{d}{2}+1\right)}\frac{1}{(-2P\cdot Y)^{\Delta}} \, ,
\end{align}
and similarly for the spinning operators. When comparing this to the dS result, one has to be careful of phases. Although the boundary point $P^2=0$ is the same, the path through which it is reached is different. In the Poincaré patch in AdS, we follow $\lambda \sim \frac{1}{z}$, and in dS, $\lambda \sim \frac{1}{(-\eta)}$, yielding a phase difference. This is precisely the phase factor we  accounted for previously when rewriting the field insertions of dS in terms of a mixture of EAdS fields. The rules previously explained suffice to fix the continuation of  the external lines as well.

\subsection{Feynman Rules for dS Amplitudes}\label{sec:rulest}

Through the careful study of the propagators in de Sitter space, we have re-derived the rules of \cite{Sleight:2021plv} to write tensorial amplitudes using effective AdS processes.\footnote{Our formulation differs slightly as we do not include possible phase factors due to derivative interactions, and we do not write the rules in Mellin Space.} This set of rules will be completed by the equivalent ones for fermion lines in sec. \ref{sec:ess} .  The transcription scheme for tensorial processes is :

\begin{itemize}
	\item First, draw the diagram in dS, specifying the dimensions of each external line. 
	\item For each internal line with $\Delta,J$, sum over two diagrams: 
	\begin{align}
		\includegraphics[valign=c,width=3cm]{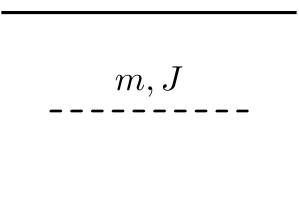}=\frac{1}{c_{\Delta_+}}\includegraphics[valign=c,width=3cm]{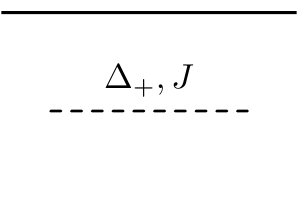}+\frac{1}{c_{\Delta_-}}\includegraphics[valign=c,width=3cm]{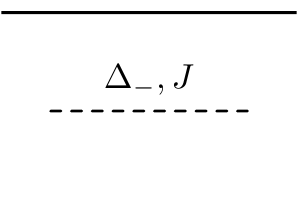} \, .
	\end{align}
	\item For each vertex, we have to fix a choice of contour. Then, we multiply by the required phases and normalisation factor to transform it into an AdS vertex. We then sum over each contour. Doing this sum explicitly for a general vertex we find : 
	\begin{align}
		\includegraphics[valign=c,width=4.5cm]{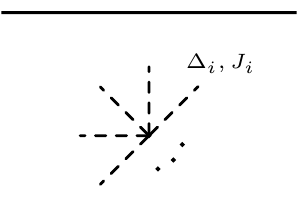}=\lambda_{(\Delta_i,J_i)}\includegraphics[valign=c,width=4cm]{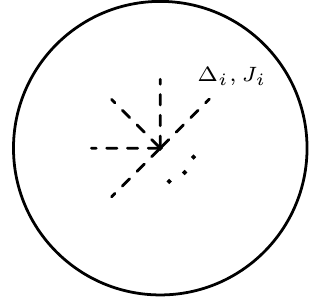}\, .
	\end{align}
	\item Finally, evaluate the diagrams using whichever methods one prefers.
\end{itemize}

For each internal lines joining two vertices, the factors of $c_{\Delta}$ get repeated on both vertices, and so we must divide the internal lines by it to avoid overcounting. The different constant we defined are

\begin{align}
	\lambda_{(\Delta_i,J_i)}&=2 \sin(\frac{\pi}{2}(d+ \sum(\Delta_i+J_i))) \prod_{i}c_{\Delta_i} \, , & c_{\Delta} &= \frac{1}{2\sin(\frac{\pi}{2}(d-2\Delta))} \, .
\end{align}

\section{Fermionic Cosmological Correlator}

We have reviewed how to rewrite a generic perturbative diagram involving tensorial operators in dS as an effective EAdS theory. The present section contains the main new result of this paper, which is an extension of the previous analysis to fermions. Since continuation of the integrals is left unchanged, the only issue left is to find how to transform the propagators. 

This is a subtle matter, owing to fermion's sensitivity to signature. There are also multiple angles of attacks, as one can consider the problem in embedding and real space, and in both cases one must heed the effect of the Wick rotation. As continuation of the scalar functions found in the propagators themselves is straightforward from the method previously used for tensors, the technicality lies in the kinematical part. 

We start by giving a parallel exposition of spinors in both dS \cite{Pethybridge:2021rwf} and EAdS \cite{Henningson:1998cd,Kawano:1999au,Nishida:2018opl}, spelling out the embedding and real-space picture in both. This sets out our convention and clarifies the different objects to be related. We then turn to the analytical continuation proper. As in the tensorial case, we use expression of the embedding object in terms of real-space one to transform them from one space to the other. We give explicit expressions for the structure transformations, which with the change of the the scalar functions in the propagators, gives the final relations. The output of this computation is presented as a set of Feynman rules incorporating fermion lines. We finish with an illustrative computation using the formalism outlined. As we make use of spinorial harmonic functions, we give back some details regarding their properties in appendix \ref{app:harmonic}.

\subsection{Fermions in de Sitter}

\begin{figure}[h]
	\centering
	\includegraphics[width=0.9\linewidth]{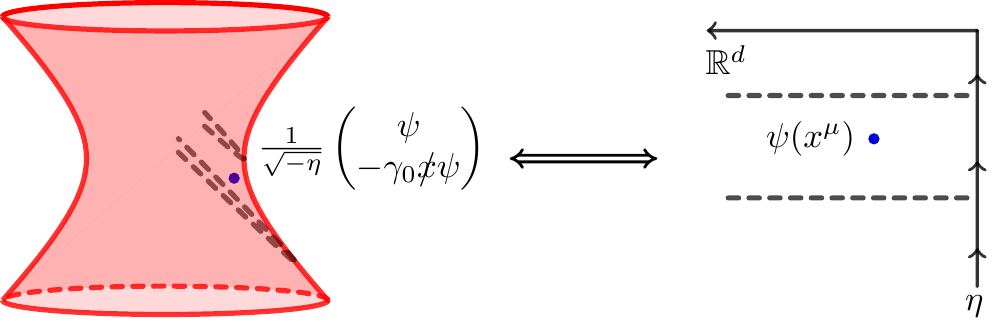}
	\caption{Uplift and Projection from the Poincaré patch of dS$_{d+1}$ into the embedding picture for the spinor bundle}
	\label{fig:spin1}
\end{figure}

We look at Dirac fermions in $\mathbb{R}^{1,d+1}$. If $d+2$ is odd, pick a direct sum of two Dirac fermions transforming in representation with the reflected gamma matrices. This allows to work in a dimension agnostic way in the embedding. We define the gamma matrices, 
\begin{align}\label{eq:gamma}
	\begin{aligned}
	 \Gamma_0 &=  \begin{pmatrix} 0 && 1 \\ -1 && 0 \end{pmatrix}\otimes\mathds{1}\, ,  \\ 
	 \Gamma_\star &=   \begin{pmatrix} 1 && 0 \\ 0 && -1 \end{pmatrix}\otimes-i\gamma_0\, ,
	\end{aligned} &&\begin{aligned}
	\Gamma_a &=\begin{pmatrix} 1 && 0 \\ 0 && -1 \end{pmatrix}\otimes \gamma_0\gamma_a \, , \\
	\Gamma_+ &=  \begin{pmatrix} 0 && 1 \\ 0 && 0 \end{pmatrix}\otimes\mathds{1} \, ,
	\end{aligned} && \begin{aligned}
	\Gamma_{d+1} &= \begin{pmatrix} 0 && 1 \\ 1 && 0 \end{pmatrix} \otimes \mathds{1}\, , \\ 
	\Gamma_- &=\begin{pmatrix} 0 && 0 \\ -1 && 0 \end{pmatrix}\otimes \mathds{1} \, .
	\end{aligned}
\end{align}
Note that even if $d+2$ is odd, $\Gamma_\star$ exists still as the representation is reducible. The real-space gamma matrices obey $\gamma_\mu^{\dagger}=\gamma_0 \gamma_\mu \gamma_0$ and $\{\gamma_\mu,\gamma_\nu\}=2\eta_{\mu\nu}$ with mostly plus metric. In the embedding we define the conjugate of $\Psi$ to be $\overline{\Psi}=\Psi^{\dagger}(-i\Gamma_0)$. In real space, we define $\overline{\psi}=i\psi^{\dagger}\gamma_0$. The minus-sign difference is there for convenience, and does not influence the reality property of bilinears \cite{VanProeyen:1999ni}. 

There are two equivalent pictures one can use. We can work with constrained embedding spinors, which are in one-to-one map with real-space spinors, as is illustrated in fig.\ref{fig:spin1}. They obey the equation 
\begin{align}
	X^{A}\Gamma_A \Psi(X)& = \Psi(X) \, ,
\end{align}
and their upper components transform like $\psi/\sqrt{-\eta}$, with $\psi$ transforming as a Dirac spinor field on $dS_{d+1}$, at the point $x^{\mu}$ associated to $X^{A}$, as in \eqref{eq:poincare} \cite{Weinberg:2010ws,Pethybridge:2021rwf}. The second perspective is to work with an unconstrained spinors $\Psi$, and contract it with a (commuting) polarisation spinor $\overline{S}$ which satisfies the constraint, projecting out the supplementary degrees of freedom. This mimics the approach used for tensors, where we restricted the polarisations to be transverse. We will proceed with this perspective now, as it gives the more succint expressions. The polarisation spinors can be solved for in terms of real-space objects
\begin{equation}\label{eq:pol1}
\begin{aligned}
	X^A \Gamma_A S = -S  \hspace{1cm}&\implies\hspace{1cm} S = \begin{pmatrix} \gamma_0 \\ -\slashed{x} \end{pmatrix}\frac{s}{\sqrt{-\eta}} \, , \\
     \overline{S}X^A \Gamma_A = \overline{S}\hspace{1cm}&\implies\hspace{1cm} \overline{S} = \frac{\overline{s}}{\sqrt{-\eta}}\begin{pmatrix} -\slashed{x}\gamma_{0} && 1 \end{pmatrix}  \, .
\end{aligned}
\end{equation}

Like the covariant derivative, one can uplift the Dirac operator to the embedding, becoming 
\begin{align}
	\slashed{\nabla} &= \slashed{\partial} - \slashed{X}\left(X\cdot \partial + \frac{d+1}{2}\right) \, .
\end{align}
This expression can be derived through a formal study of spinor bundle \cite{Trautman:1992,Trautman:1995fr}, or by noticing it is fixed by its anticommutation with the constraint. There is a slight subtlety in embedding, which is understood by looking at the following object for constrained spinors 
\begin{align}
	\overline{\Psi}\slashed{\nabla}\Psi &= -2\overline{\psi}\slashed{\nabla} \psi \, ,  & \overline{\Psi}i \Gamma_\star\Psi &= -2\overline{\psi}\psi \, .
\end{align}
From which it follows that one can equivalently write the Dirac equation as 
\begin{align}
	(\slashed{\nabla}+m)\psi = 0 \, \Leftrightarrow \, (i\Gamma_\star \slashed{\nabla}-m)\Psi =0 \, .
\end{align}
This machinery allows to solve the Dirac equation on the two-point function easily. First, note that the Wightman function is necessarily of the form 
\begin{equation}
\begin{aligned}
	\expval{\Psi(X,\overline{S}_1)\overline{\Psi}(Y,S_2)}&= \overline{S}_1 S_2 g_+ (u) + \overline{S}_1 \Gamma_\star S_2 g_- (u)  \\
	&= W_{1/2,\Delta}(X,\overline{S}_1;Y,S_2) \\
	&= \overline{s}_1W_{1/2,\Delta}(x;y)s_2 \, .
\end{aligned}
\end{equation}
The different structures take the real-space form 
\begin{align}\label{eq:structures}
	\overline{S}_1 S_2 &= \overline{s}_1\frac{\gamma^{\mu}(x-y)_\mu}{\sqrt{\eta_x \eta_y}}s_2  \, ,  & \overline{S}_1 \Gamma_\star S_2 &= \overline{s}_1\frac{i \gamma_0\gamma^{\mu}(\tilde{x}-y)_\mu}{\sqrt{\eta_x \eta_y}}s_2 \, ,
\end{align}
with $\tilde{x}^{\mu} = (-x^0,x^a)$ the time-reversal of $x^{\mu}$, $s$ and $\overline{s}$ the real-space polarisation spinors. The Wightman function then solves the Dirac equation with Bunch-Davies and Hadamard conditions. 

As before, we restrict the singularities to coincident point. For the normalisation, we take the short-distance limit of this singularity compared with the flat-space one. In dS, the geodesic distance for nearly coincident point is given by $D^2 \sim 4(1-u)$, and we should have a coincident limit $u\rightarrow 1$ of the propagator with behaviour $-\kappa_f \frac{\gamma_\mu \partial^\mu (D^2)}{2(D^2)^{\frac{d+1}{2}}}$, with $\kappa_f$ the usual fermion 2pt. function normalisation. An equivalent computation is to note that the equations of motions are satisfied everywhere except at the coincident limit $x\sim y$. Hence we can approximate the correlator by its leading behaviour in the coincident limit, and act on it with the Dirac operator. We recognise the resulting expression as a specialisation of 
\begin{align}
	\delta^{d+1}(x-y)=\frac{1}{S_{d+1}}\partial_\mu \left(\frac{x^{\mu}-y^{\mu}}{\abs{x-y}^{d+1}}\right) \, ,
\end{align}
whose normalisation we match with the one for a Lorentzian Green function.

The output of this computation is 
\begin{equation}
\begin{aligned}\label{eq:W12}
	g_{+}(u)&=\kappa \,_2 F_1\left(\Delta+\frac{1}{2},d-\Delta+\frac{1}{2};\frac{d+1}{2}; u\right)\, , \\
	g_{-}(u)&=\kappa\frac{(d-2\Delta)}{d+1}\,_2 F_1\left(\Delta+\frac{1}{2},d-\Delta+\frac{1}{2};\frac{d+1}{2}+1; u\right) \, , \\
	\kappa &=\frac{\Gamma(\Delta+\frac{1}{2})\Gamma(d-\Delta+\frac{1}{2})}{2(4\pi)^{\frac{d+1}{2}}\Gamma(\frac{d+1}{2})} \, ,
\end{aligned}
\end{equation}
with the interplay between the Dirac and Casimir equation fixing $\Delta=\frac{d}{2}+i m$. This result can also be derived from a more involved mode resummation in canonical quantisation, as in sec. \ref{sec:resum} . The different propagators follow from the previous prescription.

We will write $W_{1/2,\Delta}$ to represent interchangeably the propagators in the real or embedding space picture. The embedding space picture is the most natural one, we find, to express the kinematic transformation induced by Wick rotation. Because of this we will first perform the continuation of each element in the embedding picture independently, but then write out the result in real-space. The position space picture is still practical, as one encounters in diagrams objects of the form
\begin{align}
	\ldots W_{1/2,\Delta}(x;y) W_{1/2,\Delta}(y;z) \ldots  \, ,
\end{align}
Which are to be understood as a matrix multiplication. But in embedding space, the equivalent expression is more convoluted, as one must project out the unphysical component of the spinor indices by inserting $1+i \slashed{X}$ for contractions at position $X^{A}$. Which object we are manipulating should be clear from context.

Finally, we can take the boundary limit of the correlator if needed. One has to be careful, as the polarisation spinors have a non-trivial scaling as one goes towards the boundary, making the boundary spinor scale like $\Delta+\frac{1}{2}$ instead of $\Delta$, as is usual in the embedding picture for CFT \cite{Weinberg:2010ws}. We define the boundary polarisation spinors to be 
\begin{equation}
\begin{aligned}
	P^A \Gamma_A S_{\partial} = 0  \hspace{1cm}&\implies\hspace{1cm} S_\partial = \begin{pmatrix} \gamma_0 \\ -\gamma_i x^{i} \end{pmatrix}s \, , \\
     \overline{S}_{\partial}P^A \Gamma_A = 0 \hspace{1cm}&\implies\hspace{1cm} \overline{S}_{\partial} = \overline{s}\begin{pmatrix} -\gamma_ i x^{i}\gamma_{0} && 1 \end{pmatrix}  \, .
\end{aligned}
\end{equation}
And when we take the boundary limit of the bulk to bulk correlator, one obtains two different types of bulk-to-boundary correlator. They can be written using

\begin{equation}
\begin{aligned}
	K_{1/2,\Delta}^{dS,\pm}(X,\overline{S}_1;Y,S_2)&=\frac{N_{\Delta}}{2 \cos(\frac{\pi (d - 2 \Delta)}{2})}\frac{\overline{S}_1(1\pm \Gamma_\star)S_2}{(-2 X\cdot Y)^{\Delta+\frac{1}{2}}} \, , \\
	 N_{\Delta} &= \frac{\Gamma \left(\Delta +\frac{1}{2}\right)}{2\sqrt{\pi}^d\Gamma \left(\Delta -\frac{d-1}{2}\right)} \, .
\end{aligned}
\end{equation}

Where either $X$ or $Y$ are on the lightcone, and respectively $\overline{S}_1$ or $S_2$ are of the boundary type. The two specific propagators encountered are $K^{dS,\pm}_{1/2,\Delta_{\pm}}$, with $\Delta_{\pm} = \frac{d}{2}\pm i m$. 

\subsection{Fermions in Anti-de Sitter}

This section reproduces the previous rigmarole but in Euclidean anti-de Sitter space this time. Our approach is inspired by previous exposition of \cite{Pethybridge:2021rwf}, and draws on \cite{Nishida:2018opl}, but our conventions differ vastly from the latter. Since we are in an Euclidean space, there is no preferred reality structure to be imposed a-priori on the fermionic fields. In effect, we will use the one inherited from the de Sitter space, which in the usual picture gives a non-unitary, non-reflection positive theory in AdS. 

\begin{figure}[h]
	\centering
	\includegraphics[width=0.9\linewidth]{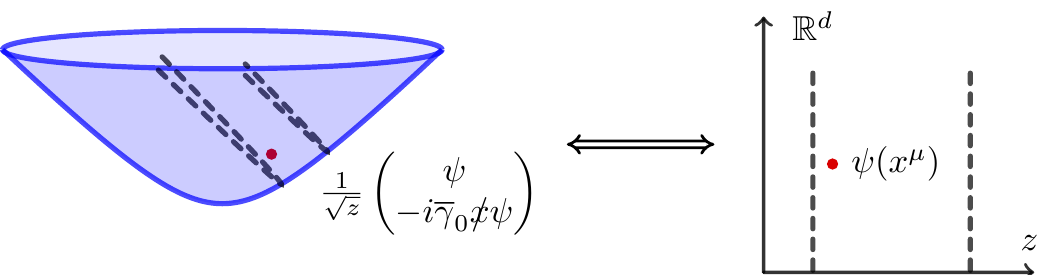}
	\caption{Uplift and Projection from the Poincaré patch of EAdS$_{d+1}$ into the embedding picture for the spinor bundle}
	\label{fig:spin2}
\end{figure}

We reuse the previous choice of gamma matrices \eqref{eq:gamma}, but now define $\overline{\gamma}_0 = -i\gamma_0$. Because in AdS $X^{A}X_A = -1$, the eigenvalue equation satisfied by the constrained embedding spinors is different. Namely, one finds 
\begin{align}
	X^{A}\Gamma_A \Psi =-i \Psi  \, ,
\end{align}
and because now the eigenvalue is complex, it follows that the conjugate satisfies the same equation with eigenvalue $-i$ as well. Such a spinor has an upper component which transforms like $\psi/\sqrt{z}$, with $\psi$ a Dirac spinor field in EAdS$_{d+1}$. This is illustrated in fig.\ref{fig:spin2}. As we did previously, we can use the associated unconstrained picture instead. We introduce polarisation spinors, which satisfy 
\begin{equation}\label{eq:pol2}
\begin{aligned}
	X_H^A \Gamma_A S_H = -iS_H  \hspace{1cm}&\implies\hspace{1cm} S_H = \begin{pmatrix} 1 \\ -i\overline{\gamma}_0\slashed{x} \end{pmatrix}\frac{s}{\sqrt{z}} \, , \\
     \overline{S}_H X_H^A \Gamma_A = -i\overline{S}_H\hspace{1cm}&\implies\hspace{1cm} \overline{S}_H = \frac{\overline{s}}{\sqrt{z}}\begin{pmatrix} \slashed{x}\overline{\gamma}_{0} && i \end{pmatrix}  \, .
\end{aligned}
\end{equation}
Note that by writing $\gamma^\mu x_\mu =\slashed{x}$ we now mean $\gamma^ix_i+\overline{\gamma}_0 z$. Which one is meant should be clear from context. Because these spinors are Euclidean, formally $s$ and $\overline{s}$ are independent quantities. We are effectively free to specify the reality structure encoded by the operation $\bullet^{\dagger}$ to make sense of it. $\overline{S}$ and $S$ are still related by $\overline{S}=S^{\dagger}(-i\Gamma_0)$, but we also have to specify how to make sense of the conjugation of the EAdS real-space spinors. We take the reality conventions imported from dS$_{d+1}$, which is not the usual choice of using the ones of Lorentzian AdS$_{d+1}$.

To find the Wightman function, we proceed as before. The Dirac operator takes the form 
\begin{align}
	\slashed{\nabla}_H &= \slashed{\partial} + \slashed{X}_H\bigg(X\cdot \partial + \frac{d+1}{2}\bigg) \, .
\end{align}
There is now a stark difference in the uplift of spinors bilinears, so we must be careful when trying to write the Dirac equation properly in the embedding. For constrained embedding spinors, one has 
\begin{align}
	\overline{\Psi}_H\Gamma_\star\slashed{\nabla}_H \Psi_H &= 2i \overline{\psi}\slashed{\nabla} \psi \, , & \overline{\Psi}_H \Psi_{H} &= 2\overline{\psi}\psi \, .
\end{align}
It follows that the Dirac equation takes exactly the same form as in de Sitter,
\begin{align}
	(\slashed{\nabla}+m)\psi = 0 \, \Leftrightarrow \, (i\Gamma_\star \slashed{\nabla}_H-m)\Psi_H =0 \, . 
\end{align}
Note however that because the polarisation spinors are quite different, the structures appearing in the propagator, although seemingly identical in the embedding, differ quite a lot once projected in real space 
\begin{align}
	\overline{S}_H^1 \Gamma_\star S_H^2 &=\overline{s}_1\frac{\gamma^\mu(x-y)_\mu}{\sqrt{z_1 z_2}}s_2 \, , & \overline{S}^1_{H} S_{H}^2 &=-\overline{s}_1\frac{\overline{\gamma}_0\gamma^\mu(\tilde{x}-y)_\mu}{\sqrt{z_1 z_2}}s_2  \, .
\end{align}

We have all the tools needed. Acting on the ansatz with the Dirac equation, one finds two independent solutions. We write the general solution as 
\begin{equation}
\begin{aligned}
	\expval{\Psi_H(X,\overline{S}_1)\overline{\Psi}_H(Y,S_2)}&= \Pi_{1/2,\Delta}^{\pm}(u)=\overline{s}_1 \Pi^{\pm}_{1/2,\Delta}(x;y)s_2 \\
	&= \overline{S}_H^1 \Gamma_\star S_H^2 f_{+}(u)\pm \overline{S}_H^1 S_H^2 f_{-}(u) \, .
\end{aligned}	
\end{equation}
The functions $f_{\pm}$ are specified through 
\begin{equation}
\begin{aligned}
	f_{+}(u)&=-\frac{\Gamma\left(\Delta +\frac{1}{2}\right)}{2\sqrt{\pi}^d \Gamma \left(\Delta-\frac{d-1}{2}\right)}\frac{1}{(-4z)^{\Delta+\frac{1}{2}}}\, _2F_1\left(\Delta +\frac{1}{2},\Delta -\frac{d}{2};2 \Delta-d +1;\frac{1}{u}\right)  \, ,\\
	f_{-}(u)&=-\frac{\Gamma\left(\Delta +\frac{1}{2}\right)}{2\sqrt{\pi}^d \Gamma \left(\Delta-\frac{d-1}{2}\right)}\frac{1}{(-4z)^{\Delta+\frac{1}{2}}}\, _2F_1\left(\Delta+\frac{1}{2},1+\Delta -\frac{d}{2};2 \Delta-d +1;\frac{1}{u}\right) \, .
\end{aligned}
\end{equation}
These functions are normalised by looking at the coincident limit and requiring them to be Euclidean Green function which have a $-\delta(\ldots)$ source term.

 The Dirac equation has two independent solutions given by $\Pi^{\pm}_{1/2,\frac{d}{2}\pm m}(u)$ for $m>0$. $m \in \mathbb{R}^{+}$ is equivalent to a unitary field in AdS in the so-called normal and alternate quantisation \cite{Henningson:1998cd,Henneaux:1998ch}. These functions have an interpretation in term of boundary conformal blocks of a fermionic bCFT, and the structures are likewise related under a Weyl transformation \cite{Herzog:2022jlx}. The principal series, though unitary in de Sitter space, can be realised in AdS through an imaginary mass for the fermion, which must be approached from the positive real part part of the complex plane. Note that this is also the type of mass which is consistent with our reality assignment for $\overline{\psi}\psi$, although unusual from an AdS perspective. This is not an issue whatsoever, because the AdS space we consider is an abstract object, a useful computational tool. 

As before, one can take the boundary limit of this object. The bulk-to-boundary correlator is easily found to be given by 
\begin{align}
	K_{1/2,\Delta}^{\pm}(X,\overline{S}_1;Y, S_{2}) &= \overline{s}_1 K_{1/2,\Delta}^{\pm}(x;y)s_2 = -N_{\Delta}\frac{\overline{S}_1(\Gamma_\star \pm 1 )S_{2}}{(-2 X \cdot Y)^{\Delta+\frac{1}{2}}} \, ,
\end{align}
where either $X$ or $Y$ are on the lightcone, and  one must replace the corresponding polarisation by the boundary one. The structures takes the explicit form 
\begin{equation}
\begin{aligned}
	\overline{S}_1(\Gamma_\star \pm 1 )S_{2,\partial} = \overline{s}_1\frac{\gamma\cdot(x-y)(1\pm \overline{\gamma}_0)}{\sqrt{z_1}}s_2 \, ,\\
	\overline{S}_{1,\partial}(\Gamma_\star \pm 1 )S_{2} = \overline{s}_1\frac{(1\mp \overline{\gamma}_0)\gamma\cdot(x-y)}{\sqrt{z_2}}s_2 \, .
\end{aligned}	
\end{equation}

To compute exchange diagrams, it is convenient to define the harmonic function \cite{Costa:2014kfa,Nishida:2018opl}, which is bilocal in the bulk and is constructed using an integrated product of bulk-to-boundary propagators with specific dimensions, 
\begin{equation}
\begin{aligned}
	\Omega_{1/2,\nu}(X, \overline{S}_1;Y,S_2)&= \frac{i}{2\pi} \int_{\partial H} [dP] \,  K_{1/2,\frac{d}{2}+i\nu}^{+}(X,\overline{S}_1;P) \slashed{P} K_{1/2,\frac{d}{2}-i\nu}^{-}(P;Y,S_2) \\
	&=\overline{s}_1 \Omega_{1/2,\nu}(x;y) s_2 \, .
\end{aligned}
\end{equation}
Where by $[dP]$ we denote the conformally invariant integral over the light-cone \cite{Simmons-Duffin:2014wb}. Through an explicit computation, easily done in the embedding using the formula from appendix A of \cite{Herzog:2022jlx}, one can relate it to the functions previously found 
\begin{equation}
	\Omega_{1/2,\nu}(x;y) =\frac{1}{2\pi}\left(\Pi^{+}_{1/2,h+i\nu}(x;y)-\Pi^{-}_{1/2,h-i\nu}(x;y)\right)  \, .
\end{equation}
The harmonic function satisfies the differential equation 
\begin{align}
	(\slashed{\nabla}_{x}+i\nu)\Omega_{1/2,\nu}(x;y) = 0 = \Omega_{1/2,\nu}(x;y) (\overset{\leftarrow}{\slashed{\nabla}}_y-i\nu) \, ,
\end{align}
making them an orthonormal basis. We give in appendix \ref{app:harmonic} a detailed derivation of the orthogonality relation 
\begin{align}
	\int_{AdS} d^{d+1} y  \, \Omega_{1/2,\nu}(x;y)\Omega_{1/2,\nu'}(y;z)= \delta(\nu-\nu')\Omega_{1/2,\nu}(x;z) \, , 
\end{align}
which explains the specific normalisation chosen for the integral. This type of relation allows us to rewrite the propagators we found in terms of a harmonic decomposition
\begin{align}\label{eq:spinharm}
	\Pi_{1/2,h\pm i\nu}^{\pm}(x;y)=-i\int_{-\infty}^{\infty}\frac{d\nu'}{\nu'-\nu\pm i\epsilon}\Omega_{1/2,\nu'}(x;y) \, .
\end{align}
These expressions are particularly symmetric because one has to analytically continue the result with real $m>0$ to $i\nu$, and the $\epsilon$ part is necessary to pick up the right contour. Away from the principal series, the two propagators have more distinct expressions.

\subsection{Analytic Continuation}

Having properly defined the different players, we can consider their relation under the analytic continuation  \ref{eq:cont} . The fermions have a more involved kinematic behaviour under this transformation, as we can illustrate quickly through a simple argument.

We have seen that the embedding coordinates are deformed along the path $X\rightarrow e^{\mp i\frac{\pi}{2}}X$. At the level of the eigenvalue equation satisfied by the constrained spinors, this gives a are mapping from one to the other, up to a sign. This  sign is meaningful, since we pick spinors in EAdS to obey the specific eigenvalue equation $X^{A}_{H}\Gamma_A \Psi_{H}(X)=-i\Psi_H(X)$, from which it follows that $\Psi^{r}\rightarrow \Psi_{H}$, while $\Psi^{l}\rightarrow \Gamma_\star \Psi_{H}$. Likewise, since $\overline{\Psi}$ has opposite eigenvalue, it follows that one has $\overline{\Psi}^{l}\rightarrow \overline{\Psi}_{H}$ and $\overline{\Psi}^{r}\rightarrow \Gamma_\star\overline{\Psi}_{H}$. This implies the Wick rotation of the spinors act non-trivially on the kinematical structures entering the Wightman function. 

To be more quantitative about this behaviour, we can turn to the unconstrained-spinor contracted with polarisation picture. We use the explicit expressions \eqref{eq:pol1},\eqref{eq:pol2} and perform the Wick rotation of \eqref{eq:cont}, to obtain the relations
\begin{align}
	\begin{aligned}
		\overline{S}^{l}[\overline{s}]&\rightarrow +e^{+i\frac{\pi}{4}}\overline{S}_H[\overline{s}\overline{\gamma}_0]\Gamma_\star \, , \\
		S^{l}[s]&\rightarrow -e^{+i\frac{\pi}{4}} S_H[\overline{\gamma}_0 s] \, ,
	\end{aligned} && 
	\begin{aligned}
		\overline{S}^{r}[\overline{s}]&\rightarrow -e^{-i\frac{\pi}{4}}\overline{S}_H[\overline{s}] \, , \\
		S^{r}[s]&\rightarrow -e^{-i\frac{\pi}{4}}\Gamma_\star S_H[s] \, .
	\end{aligned}
\end{align}
By combining this result with the hypergeometric identities \eqref{eq:id1}, \eqref{eq:id2}, we can analytically continue the propagators. Replacing the structures and the functions multiplying them by their analytically continued expressions, we decompose the result by inspection into a sum of AdS propagators.

Consider first the simplest case of fields in the $lr$ and $rl$ configuration,
\begin{equation}
\begin{aligned}
	W_{1/2,\Delta}^{lr}(x;y)& \rightarrow\frac{-\overline{\gamma}_0}{2\cos(\frac{\pi(d-2\Delta)}{2})}\left(\Pi^{+}_{1/2,\Delta}(x;y)-\Pi^{-}_{1/2,d-\Delta}(x;y)\right) \, , \\
	W_{1/2,\Delta}^{rl}(x;y)&\rightarrow\left(\Pi^{+}_{1/2,\Delta}(x;y)-\Pi^{-}_{1/2,d-\Delta}(x;y)\right)\frac{\overline{\gamma}_0}{2\cos(\frac{\pi(d-2\Delta)}{2})} \, .
\end{aligned}
\end{equation}
The non-trivial factors are now due to the spinor polarisation. Looking at fields in the same configuration,
\begin{equation}
\begin{aligned}
	W_{1/2,\Delta}^{ll}(x;y)&\rightarrow\frac{-\overline{\gamma}_0}{2\cos(\frac{\pi(d-2\Delta)}{2})}\left(e^{i\pi \Delta}\Pi^{+}_{1/2,\Delta}(x;y)+e^{i\pi(d-\Delta)}\Pi^{-}_{1/2,\Delta}(x;y) \right)\overline{\gamma}_0  \, .\\
	W_{1/2,\Delta}^{rr}(x;y)&\rightarrow\frac{1}{2\cos(\frac{\pi(d-2\Delta)}{2})}\left(e^{-i\pi \Delta}\Pi^{+}_{1/2,\Delta}(x;y)+e^{-i\pi(d-\Delta)}\Pi^{-}_{1/2,\Delta}(x;y) \right) \, .
\end{aligned}
\end{equation}

Like we did for tensors, these different propagators decomposition can be reformulated in terms of a change of basis to a pair of shadow-dual AdS fields 
\begin{align}\label{eq:chang}
	\begin{aligned}
		\psi^{l}&= -\overline{\gamma}_0 \left(e^{i\frac{\pi}{2}\Delta}\psi_{+}+e^{i\frac{\pi}{2}(d-\Delta)}\psi_{-}\right) \, , \\
		\psi^{r}&= \left(e^{-i\frac{\pi}{2}\Delta}\psi_{+}-e^{-i\frac{\pi}{2}(d-\Delta)}\psi_{-}\right) \, ,
	\end{aligned} && 
	\begin{aligned}
		\overline{\psi}^{l}&= \left(e^{i\frac{\pi}{2}\Delta}\overline{\psi}_{+}+e^{i\frac{\pi}{2}(d-\Delta)}\overline{\psi}_{-}\right)\overline{\gamma}_0 \, , \\ 
		\overline{\psi}^{r}&= \left(e^{-i\frac{\pi}{2}\Delta}\overline{\psi}_{+}-e^{-i\frac{\pi}{2}(d-\Delta)}\overline{\psi}_{-}\right)  \,,
	\end{aligned}
\end{align}
whose propagators are normalised in a unusual way 
\begin{align}
	\expval{\psi_{+}(x)\overline{\psi}_{+}(y)}&=\frac{\Pi_{1/2,\Delta}^{+}(x;y)}{2\cos(\frac{\pi(d-2\Delta)}{2})}\, , & \expval{\psi_{-}(x)\overline{\psi}_{-}(y)}&=\frac{\Pi_{1/2,d-\Delta}^{-}(x;y)}{2\cos(\frac{\pi(d-2\Delta)}{2})} \, .
\end{align}
Note that these equations are consistent with the eigenvalue discussion. The pair of AdS fields obtained sit in the two quantisation schemes of spinors with imaginary masses $\pm im$. It is interesting to note how  for tensors the ``spinning" part of the phase factor, $e^{\pm i\frac{\pi}{2}J}$, came about because of the analytical continuation of the tensor structures, which was quite simple, while here we have a more intricate phase structure appearing. Note also how the bosonic correlators where normalised using a $\sin(\ldots)$, while the fermionic ones uses a $\cos(\ldots)$ factor. Using the relations \eqref{eq:chang} one can discuss the analytical continuation of general interaction terms.

As in the bosonic case, these replacement rules are enough to take into account the change in the bulk-to-boundary propagators as well. These fully specifies the rules to translate perturbation theory when fermions are involved. They are enough to perform a practical computation involving spinors in expanding space-time. 

\subsection{Feynman Rules With Spinors}\label{sec:ess}

Because of the slightly more involved phase in the fermion change of basis, pairs of fermion lines alter the rules of sec.\ref{sec:rulest}. If a pair of fermion lines in different quantisation are joined together at a vertex, they pick up a further $-1$ difference between the left and right contour from the bosonic case, turning the $\sin$ into a $\cos$. 

Taking into account the previous results, we find that to evaluate a fermionic correlator, one should proceed as follows

\begin{itemize}
	\item First, draw the diagram in dS, specifying the dimensions of each external line. 
	\item For each internal spinor line, sum over two diagrams with weight $\Delta_{\pm}$, and divide with factors of $c_{f,\Delta_i}$.
	\begin{align}
		\includegraphics[valign=c,width=3cm]{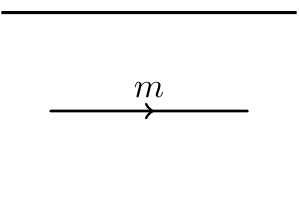}=\frac{1}{c_{f,\Delta_+}}\includegraphics[valign=c,width=3cm]{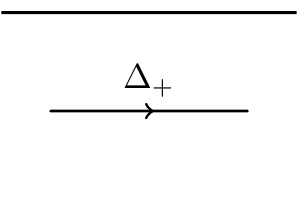}+\frac{1}{c_{f,\Delta_-}}\includegraphics[valign=c,width=3cm]{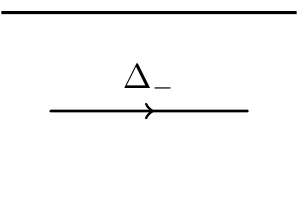} \, .
	\end{align}

	\item For each vertex, we have to fix a choice of contour. Then, We multiply by the required phases and normalisation factor to transform it into an AdS vertex. We then sum over each contour. Doing this sum explicitly for a general vertex with one pair of spinors, we find 
	\begin{align}
		\includegraphics[valign=c,width=5cm]{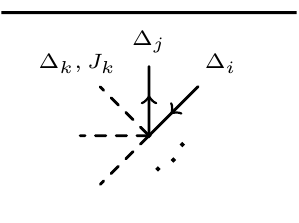}=\lambda_{(\Delta_k,J_k)}^{ij}\includegraphics[valign=c,width=4cm]{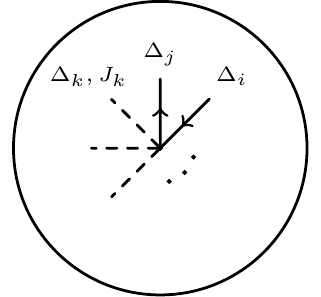}\, .
	\end{align}
	\item Finally, evaluate the AdS diagram using whichever method one prefers.
\end{itemize}

The constants are given by 
\begin{equation}
\begin{aligned}
	\lambda_{(\Delta_k,J_k)}^{(ij)}=\left( c_{f,\Delta_i}c_{f,\Delta_j} \prod_{i}c_{\Delta_i} \right) \bigg(&2\delta_{ij}\cos(\frac{\pi}{2}(d+\sum(\Delta_i+J_i))) \\
	&+2i\sigma_{ij}\sin(\frac{\pi}{2}(d+\sum(\Delta_i+J_i))) \bigg) \, ,
\end{aligned}
\end{equation}
Where in the sum one takes all the legs of the vertex, including fermions with $J=1/2$. The normalisation and structure are specified by 
\begin{align}
	c_{f,\Delta} &=c_{\Delta+\frac{1}{2}}= \frac{1}{2\cos(\frac{\pi}{2}(d-2\Delta))} \, ,  & \sigma_{ij}&=\begin{pmatrix}
		0 & 1 \\ 1 & 0 
	\end{pmatrix} \, .
\end{align}

The generalisation to more pairs of spinors insertions follows through straightforwardly using the relations \eqref{eq:chang} and choosing a specific vertex type and summing over the ordering label. There is no new feature beyond the ones already present here.

\subsection{An Example of Cosmological Correlator}\label{sec:comput}

Using the formalism we develop, we can perform an analytical computation of a cosmological correlator with fermions. To illustrate it, we will consider a slightly contrived example which has the benefit of being relatively tractable. 
 
We will be looking at the fermion-exchange process between two scalars and two spinors in different boundary conditions, interacting through a Yukawa term. Methods to study such configurations are known in AdS \cite{Kawano:1999au,Nishida:2018opl,Giombi:2021cnr}. To lighten the notation, since all the polarisation spinors which will appear are of the boundary type, we will drop the $\partial$ label in this section. We also abbreviate $h=\frac{d}{2}$ throughout. 

We use the rules previously derived to relate the tree-level process to a linear combination of Witten diagrams. For each Witten diagram, we rewrite the bulk-to-bulk spinor propagator using the harmonic representation in eq. \eqref{eq:spinharm}. We gather the expressions together, 
\begin{align}\label{eq:diag}
		\includegraphics[valign=c,width=5cm]{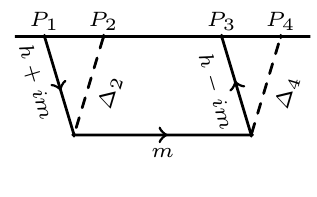}=\int d\nu \, r_{\nu} \int [dP] \, \includegraphics[valign=c,width=5cm]{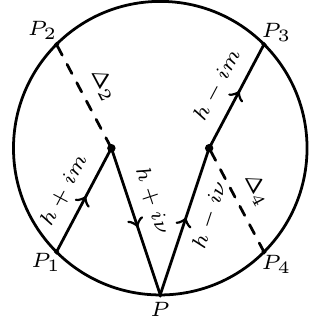} \, .
\end{align}

The precise configuration chosen is motivated by the ensueing selection rules that simplify the computation to be performed. Like in boundary CFT, the boundary spinors have a "chirality", and the structures to be sewn together by the boundary integral over $P$ are simpler in this configuration. The spectral parameter entering the integral is given by 
\begin{equation}
\begin{aligned}
	r_{\nu} = i\frac{2\lambda^2}{\pi}c_{\Delta_2} c_{\Delta_4} c_{f,h+im} &c_{f,h-im}\bigg( \frac{c_{f,h+im}\cos(\frac{\pi}{2}(2d+\Delta_4))\sin(\frac{\pi}{2}(2d+2im+\Delta_2))}{\nu-m+i\epsilon}\\
	 &+\frac{c_{f,h-im}\cos(\frac{\pi}{2}(2d+\Delta_2))\sin(\frac{\pi}{2}(2d-2im+\Delta_4))}{\nu-m-i\epsilon} \bigg) \, .
\end{aligned}
\end{equation} 

By using the harmonic function, we have already split the difficulty. We have two related bulk integrals to compute, and a final boundary one. Both bulk integrals can be computed starting from the simpler diagrams 
\begin{align}
	\includegraphics[valign=c,width=3cm]{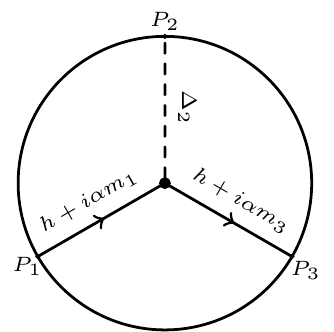}=& \int_{AdS} \frac{  DX  \, 2C_{\Delta_2}N_{h+i\alpha m_1}N_{h-i\alpha m_3} \alpha \overline{S}_1(\Gamma_{\star}+\alpha)S_3}{(-2P_1\cdot X)^{h+i\alpha m_1+\frac{1}{2}}(-2P_2\cdot X)^{\Delta_2}(-2P_3\cdot X)^{h-i\alpha m_3+\frac{1}{2}}}
\end{align}
The numerator of the integrand is found as follows. In the embedding, we free the spinor indices of the bulk-to-boundary, multiply by $1+i\slashed{X}$ to project out the unphysical components, and simplify. In real-space, we simply multiply through the gamma matrices, finding the same result once uplifted back. The configuration we chose for the 4-point function is tuned so that one can get rid of the factors of $X^{A}$ on both 3-point functions.

We have written $C_{\Delta}=N_{\Delta-\frac{1}{2}}$, the normalisation of the scalar propagator. As the remaining numerator is independent of $X^{A}$, it reduces to the known result \cite{Freedman:1998tz,Cornalba:2008qf,Costa:2014kfa} of a scalar 3-point function vertex integration in AdS with some specific weight combination. Collecting the factors, we find

\begin{align}
	n_{\Delta_2,m_1,m_3,\alpha}\frac{\alpha \overline{S}_1(\Gamma_{\star}+\alpha)S_3}{(P_{12})^{\Delta_{12,3}}(P_{23})^{\Delta_{23,1}}(P_{31})^{\Delta_{31,2}}} \, .
\end{align}

and we used the abbreviation $P_{ij}=-2P_i \cdot P_j$, $\Delta_{ij,k}=\frac{\Delta_{i}+\Delta_j-\Delta_k}{2}$, and one should set $\Delta_1=h+\frac{1}{2}+i\alpha m_1$ and $\Delta_3=h+\frac{1}{2}+i\alpha m_3$. The coefficient is given by
\begin{align}
	n_{\Delta_2,m_1,m_3,\alpha}=2 C_{\Delta_1}^2C_{\Delta_2}^2C_{\Delta_3}^2\frac{\sqrt{\pi}^{d}\Gamma(\frac{\Delta_1+\Delta_2+\Delta_3-d}{2})\Gamma(\Delta_{12,3})\Gamma(\Delta_{23,1})\Gamma(\Delta_{31,2})}{2\Gamma(\Delta_1)\Gamma(\Delta_2)\Gamma(\Delta_3)} \, .
\end{align}

With the abbreviations of earlier still applying. This factor, though unwieldy, will simply become a part of the spectral integral left-over at the end. 

We can now rewrite the left-hand side of \eqref{eq:diag} as
\begin{align}
	\includegraphics[valign=c,width=5cm]{diagram/exchange1}&=\int d\nu \mathcal{R}_{\nu} \mathcal{W}^{(h+im,\Delta_2,h-im,\Delta_4)}_{\nu}(P_1,P_2,P_3,P_4) \, .
\end{align}
The spectral function $\mathcal{R}_{\nu}$ is given by 
\begin{align}
	\mathcal{R}_{\nu}&=-r_{\nu}n_{\Delta_2,m,\nu,+}n_{\Delta_4,\nu,m,-} \, ,
\end{align}
and the function $\mathcal{W}_{\nu}$ is the (non-properly normalised) Euclidean conformal partial waves associated to the exchange of a primary of spin $1/2$. There are multiple such partial waves in general, but this corresponds to the single-one allowed by the selection rules of this ``chirality'' configuration. It is found by recognising the leftover integral over $P$ as a conformal integral \cite{Ferrara:1972kab,Dolan:2001wg,Dolan:2004up,Dolan:2012wt} written in the embedding space \cite{Simmons-Duffin:2014wb}. It can be further expressed, using Mellin space for example, or by explicit evaluation in some dimensions.

\section{Källen-Lehman Decomposition For Spinors}

Our previous result has probed the analytical structure of spinorial insertions in de Sitter. The goal of the present section is to showcase a different perspective on this issue, by investigating the notion of unitarity \cite{Bros:1990cu,Bros:1994dn,Bros:1995js,Bros:1998ik,Bros:2010rku}. The central result of this section is a proof that the Wightman function of generic spinor field in dS admits a positive spectral decomposition over the principal series.  The idea is the same as in flat space \cite{Weinberg2005}, that one can insert in the correlator a complete set of state transforming irreducibly under the isometry group. The contribution to the correlator of a specific family of state then takes the form of an integral over a product of wavefunctions. These are entirely fixed by symmetry to obey the Dirac equation, and are precisely the mode functions one finds by canonically quantising the free Dirac field. Finding the explicit spectral decomposition becomes equivalent to performing the mode resummation for the Wightman function of free fields.

Our exposition proceeds in reverse order. We first review the canonical quantisation of the free scalar field in de Sitter space, to introduce a few useful functions and tools. We then proceed to do the same for the free Dirac field, rederiving \eqref{eq:W12}. Finally, using this knowledge, we use the methods of \cite{Hogervorst:2021uvp} to fix the wavefunctions of a generic Dirac spinor over an on-shell state to be the free field modes, completing our proof. 

\subsection{Free Scalar Field, a Brief Reminder}

The quantisation of the scalar field is a standard procedure \cite{Tagirov:1972vv,Chernikov:1968zm,Schomblond:1976xc,Mottola:1984ar,Allen:1985ux,Spradlin:2001pw,Anninos:2012qw}, which we review here to set our conventions and give a springboard for the more complicated case of the Dirac spinor. 

Consider the action for a free massive scalar field,
\begin{equation}
\begin{aligned}
	S& = \frac{1}{2}\int_{dS} d^{d+1}x \,  \phi\left(\Box-m^2\right)\phi =-\frac{1}{2}\int \frac{d^{d}x \, d\eta}{(-\eta)^{d+1}} \left( \partial_\eta \phi \partial_\eta \phi-\partial_i \phi\partial^i \phi + m^2 \phi^2 \right) \, ,
\end{aligned}
\end{equation}
with equations of motions 
\begin{align}\label{eq:fund}
	\eta^2 \partial_\eta^2 \phi-(d-1)\eta \partial_\eta \phi+\left(\frac{d^2}{4}+\nu^2 -\eta^2 \nabla^2\right) \phi &= 0 \, . &  m^2 &= \frac{d^2}{4}+\nu^2  \, ,
\end{align}
This equation will reappear in another form when we solve the Dirac equation. Naturally, it simplifies in Fourier space, where one finds that the solutions to \eqref{eq:fund} are linear combinations of Hankel functions multiplied by power-laws

\begin{align}
	\phi(x,\eta) &= \int \frac{d^{d}k}{(2\pi)^{d}} \ e^{i k\cdot x}\phi_k(\eta) \, , & \phi_k(\eta) \sim (-\eta)^{\frac{d}{2}} H_{i\nu}^{(1/2)}(-k\eta) \, .
\end{align}

These functions have many interesting properties, among which that they appear generically in the modes of particles in de Sitter space. We pick the vacuum of the theory, i.e. some specific behaviour for our solution. In the Bunch-Davies vacuum, as $\eta \rightarrow -\infty$, we recover the polarisations of a massless particle of the same type in flat-spacetime, i.e. 
\begin{align}
	\lim_{k}\phi_{k}(\eta) \propto \frac{e^{-ik \eta}}{\sqrt{2k}} a_{k}+ \ldots \, .
\end{align}

It is useful to define functions which have well normalised limits already
\begin{align}\label{eq:hmode}
	h_{\alpha}(k\eta) &= \frac{\sqrt{\pi}e^{\frac{-i\alpha\pi}{2}-i\frac{\pi}{4}}}{2} H^{(2)}_{\alpha}(-k\eta) \, , & \overline{h}_{\alpha}(k\eta)&= \frac{\sqrt{\pi}e^{\frac{i\alpha \pi}{2}+i\frac{\pi}{4}}}{2} H^{(1)}_{\alpha}(-k\eta) \, .
\end{align}
These satisfy 
\begin{align}
	\overline{h}_{\overline{\alpha}}(k\eta) &= \left( h_{\alpha}(k\eta) \right)^{\dagger} \, , & h_{-\alpha}(k\eta) &= h_{\alpha}(k\eta)  \, ,& \lim_{\eta\rightarrow -\infty}h_{\alpha}(k\eta) &=\frac{1}{\sqrt{-\eta}}\frac{e^{i k \eta}}{\sqrt{2k}} \, ,
\end{align}

The most general solution for the field with the given boundary conditions is then 
\begin{align}
	\phi(x,\eta)= \int \frac{d^{d}p}{(2\pi)^{d}}  \ \underbrace{(-\eta)^{\frac{d}{2}}e^{i p \cdot x} \overline{h}_{i\nu}(p \eta)}_{f_p(\eta)} a_{p}+ (-\eta)^{\frac{d}{2}}e^{-i p \cdot x}h_{i\nu}(p\eta) a_{p}^{\dagger} \, ,
\end{align}
where we also added the negative energy modes. $f_p(\eta)$ is the wavefunction overlap between the field $\phi(x,\eta)$ and a one-particle state in the Fock space of the $a_{p}$ oscillator. 

We can check that this decomposition and normalisation reproduce the commutation relations for the fields and the creation-annihilation operator, by normalising using the Klein-Gordon inner product. Consider a cconstant $\eta$ slice, the inner product is given by 
\begin{align}
	(f,g) = i\int \frac{d^{d}x}{(-\eta)^{d-1}} (f \nabla_\eta g^{\star}-g^{\star} \nabla_\eta f)
\end{align}
The orthonormality of the wavefunctions follows directly 
\begin{equation}
\begin{aligned}
	(f_p,f_k)&=-i \int \frac{d^{d}x}{(-\eta)^{d-1}}e^{i(p-k)\cdot x} (-\eta)^{d}\left( h_{i\nu}(k\eta)\partial_\eta \overline{h}_{i \nu}- \overline{h}_{i\nu}(k\eta)\partial_\eta h_{i \nu}\right) \\
	&=i(2\pi)^{d}\delta^{d}(p-k)\left( h_{i\nu}(k\eta)\eta \partial_\eta \overline{h}_{i \nu}- \overline{h}_{i\nu}(k\eta)\eta \partial_\eta h_{i \nu}\right) \\
	&= (2\pi)^{d}\delta^{d}(p-k)
\end{aligned}	
\end{equation}

With this mode decomposition in hand, one can check that the canonical commutation relations impose that the operators $a$ and $a^{\dagger}$ be of Fock type. This is all that is needed to turn to the question of the Wightman function. By definition, it is given through the evaluation of the 2-point function on the Fock-vacuum $\ket{0}$ of the creation operators, 

\begin{equation}
\begin{aligned}
	\expval{\phi(x,\eta_1)\phi(y,\eta_2)}&=\bra{0}\phi(x,\eta_1)\phi(y,\eta_2)\ket{0} \\
	&=(\eta_1 \eta_2)^{\frac{d}{2}}\int \frac{d^{d}p \, d^{d}k}{(2\pi)^{2d}} e^{i(p\cdot x-k\cdot y)}\overline{h}_{i\nu}(p\eta_1)h_{i\nu}(k\eta_2)\bra{0}a_{p}a^{\dagger}_{k}\ket{0}\\
	&=(\eta_1 \eta_2)^{\frac{d}{2}}\int \frac{d^{d}p}{(2\pi)^{d}} \ e^{ip\cdot(x-y)} \ \overline{h}_{i\nu}(p\eta_1)h_{i\nu}(p\eta_2)\\
	&=\frac{\Gamma(\frac{d}{2}+i\nu)\Gamma(\frac{d}{2}-i\nu)}{(4\pi)^{\frac{d+1}{2}}\Gamma(\frac{d+1}{2})} \,_2 F_1\left(\frac{d}{2}+i\nu,\frac{d}{2}-i\nu;\frac{d+1}{2};u \right) = I_{\nu}(u) \, .
\end{aligned}
\end{equation}
This does reproduce the result \eqref{eq:w0}, with the variable $u$ given as in \eqref{eq:crossratio}. Performing this Fourier transform is quite involved. The interested reader can follow the steps outlined in appendix A of \cite{Bunch:1978yq}. These are the analytical continuation of an integral representation for the product of Bessel-K functions, followed by a chain of change of variables to massage the expression into the Euler parametrisation of the hypergeometric function.

\subsection{Quantising the Free Dirac Spinor}\label{sec:resum}

Armed with the results obtained for the scalar, we can turn to the more complex Dirac field. The modes in Poincaré coordinates have been considered in \cite{Cotaescu:2016sof}. Consider a free massive Dirac fermion in dS$_{d+1}$. Its action is given through
\begin{align}
	S= - \int \frac{d^{d}x \, d\eta }{\eta^{d+1}} \overline{\psi}\left(\slashed{\nabla}+m\right)\psi \, ,
\end{align}
and noting the classic result $\slashed{\nabla}=\eta \gamma^{\mu}\partial_\mu+\frac{d}{2}\gamma_0$ \cite{Henningson:1998cd} for the Dirac operator, the equation of motions are 
\begin{align}
	(\eta\partial_\eta-\frac{d}{2}+m\gamma_0+\eta \gamma_0\gamma_i \partial_i)\psi =0  \, .
\end{align}

We divide $\psi$ in $\gamma_0$ eigenvectors $\psi_+$ and $\psi_-$, with $\gamma_0 \psi_{\pm}=\pm i \psi_{\pm}$, and obtain a system of coupled differential equations
\begin{equation}\label{eq:dirac1}
\begin{aligned}
	\left(\eta\partial_\eta-\frac{d}{2}+i m \right) \psi_{+}(\eta,k)&= +\eta k^i\gamma_i \psi_{-}(\eta,k) \, , \\
	\left(\eta\partial_\eta-\frac{d}{2}-i m \right) \psi_{-}(\eta,k)&= -\eta k^i\gamma_i \psi_{+}(\eta,k)\, .
\end{aligned}
\end{equation}

At this point, it is useful to name these operators and note some of of their properties. Defining $f_{\nu}=(-\eta)^{\frac{d}{2}}h_{\nu}(k\eta)$, we have 
\begin{align}\label{eq:raise}
	L_{\pm} &= \eta \partial_\eta -\left(\frac{d}{2}\pm\nu\right) \, , & L_{\pm} f_{\nu}& = i k \eta f_{\nu \pm 1} \, .
\end{align}

To solve \eqref{eq:dirac1}, we can iterate the equations to obtain a decoupled set of 2nd order equations, for each individual eigenvector
\begin{equation}
\begin{aligned}
	\left(\eta^2 \partial_\eta^2 - d\eta \partial_\eta+ k^2 \eta^2 + \frac{(d+1)^2}{4}+\left(m \mp \frac{i}{2}\right)^2  \right) \psi_{\pm}(\eta,k) &=0 \, .
\end{aligned}
\end{equation}
These are equivalent to the equation for scalars \eqref{eq:fund}, with the modification $d\rightarrow d+1$, and identification $ i\nu_{\pm} = im \pm \frac{1}{2}$. The solutions can now be written using the functions \eqref{eq:hmode}, some $k$ dependant constants, and an orthonormal basis of eigenspinors of $\gamma_0$, $\xi_{r}$ and $\chi_{r}$. This choice is particularly convenient to analyse physically the mode expansion when $d=2k$, because one can take $\gamma_0$ diagonal and all other matrices off-diagonal, giving a simple structure. When $d=2k+1$, it is more convenient to diagonalise $\gamma_\star$, giving a chiral representations. In that case, it is then interesting to redecompose the basis into chiral eigenmodes. This is not necessary for our computation however. We will pick $\xi_{r}$ to be an orthonormal basis of the half spinor, with $\gamma_0 \xi = + i \xi$. It spans a space with dimension $\frac{1}{2}2^{[\frac{d+1}{2}]}$, which corresponds to the physical polarisations state of a spinor.

 From here, it is straightforward to work out the positive energy modes, which correspond to asymptotic behaviour $e^{-ik\eta}$, selecting out the $\overline{h}$ solution. Further imposing the first order equation relating $\psi_{+}$ and $\psi_-$, fixes the relative coefficients, and expresses $\chi$ modes in term of $\frac{k\cdot \gamma}{k}\xi_{r}$. To find the negative energy mode, we simply have to consider repeating this operation while sending $k^{i}\rightarrow -k^{i}$, and solving the analogous equation for $h$. We now instead relate the $+i$ $\gamma_0$-eigenvectors to the $-i$ ones.
  
 The end result of this computation is the mode decomposition for the spinor field
\begin{equation}\label{eq:modepsi}
\begin{aligned}
	\psi(x,\eta)&= \int \frac{d^{d}p}{(2\pi)^{d}} \sum_{r} e^{ip\cdot x} u_{r,p}(\eta) a_{r,p}+ e^{-ip\cdot x}v_{r,p}(\eta)b^{\dagger}_{r,p}  \, ,\\
	u_{r,p}(\eta)&=\frac{1}{\sqrt{2}}(-\eta)^{\frac{d+1}{2}}\left( \sqrt{p}\ \overline{h}_{im+\frac{1}{2}}(p\eta)-i \frac{p\cdot\gamma}{\sqrt{p}}\  \overline{h}_{i m-\frac{1}{2}}(p\eta)\right)\xi_{r} \, , \\
	v_{r,p}(\eta)&= \frac{1}{\sqrt{2}}(-\eta)^{\frac{d+1}{2}}\left(i\frac{p\cdot \gamma}{\sqrt{p}}\ h_{im+\frac{1}{2}}(p\eta)+\sqrt{p}\ h_{i m-\frac{1}{2}} (p\eta)\right)\chi_{r} \, .
\end{aligned}
\end{equation}
The results for the conjugate field are similarly found
\begin{equation}
\begin{aligned}
	\overline{\psi}(x,\eta)&=\int \frac{d^{d}p}{(2\pi)^{d}} \sum_{r} e^{-ip\cdot x} \overline{u}_{r,p}(\eta) a_{r,p}^{\dagger}+ e^{ip\cdot x}\overline{v}_{r,p}(\eta)b_{r,p} \, , \\
	\overline{u}_{r,p}(\eta)&=\frac{1}{\sqrt{2}}(-\eta)^{\frac{d+1}{2}}\overline{\xi}_{r}\left( \sqrt{p}\ h_{im-\frac{1}{2}}(p\eta)-i \frac{p\cdot\gamma}{\sqrt{p}}  \ h_{i m+\frac{1}{2}}(p\eta) \right) \, , \\
	\overline{v}_{r,p}(\eta)&= \frac{1}{\sqrt{2}}(-\eta)^{\frac{d+1}{2}}\overline{\chi}_{r}\left( i\frac{p\cdot\gamma}{\sqrt{p}}\ \overline{h}_{im-\frac{1}{2}}(p\eta)+\sqrt{p}\ \overline{h}_{i m+\frac{1}{2}}(p\eta)  \right)   \, .
\end{aligned}
\end{equation}
Note $\overline{\xi}_{r}\gamma_0 = i\overline{\xi}_{r}$ satisfy the same eigenvalue equation as its conjugate. We remind the reader interested in reproducing the expression for the conjugate that in our conventions $\psi \rightarrow \overline{\psi}=i\psi^{\dagger}\gamma_0$, $\gamma_0^{\dagger}=-\gamma_0$ \cite{VanProeyen:1999ni}. For future convenience, we will sometimes write these modes as $u_{r,p}^{+m}(\eta)$, to specify the sign of the mass in the Dirac equation.

The normalisation we picked for the mode can be understood through the following argument. Consider a free massless spinor in Minkowski spacetime. Its polarisations spinors will have a generic shape $\sqrt{p^{\mu}\sigma_\mu}\xi_{r} \sim \frac{p +\sigma_i p^{i}}{\sqrt{2p}}\xi_r$, which is the one reproduced by \eqref{eq:modepsi} in the early-time limit. One does not have to rely on this argument however, since these modes can be orthonormalised with respect to the Dirac inner product, like in bosonic case, yielding back this normalisation.

The Dirac inner product is given through the integral over a space-like slice of the conserved Dirac current. A quick way to rederive it is as follows. We write a manifestly vanishing integral over a volume $\Omega$ between two times slices $\Sigma_i$ and $\Sigma_f$, 
\begin{equation}
\begin{aligned}
	0&=i\int_{\Omega} \frac{d^{d}x\,  d\eta}{(-\eta)^{d+1}}\left(\overline{\psi}_1 (\slashed{\nabla} -m)\psi_2-\overline{\psi}_1(-\overset{\leftarrow}{\slashed{\nabla}}-m)\psi_2  \right) \\
	&=i\int d^{d}x \, d\eta \, \partial_\eta \left( \frac{\overline{\psi}_1\gamma^0\psi_2}{(-\eta)^{d}}\right)=i\int_{\Sigma_i-\Sigma_f}\frac{d^{d}x}{(-\eta)^{d}}\overline{\psi_1}\gamma^{0}\psi_2  \, ,
\end{aligned}
\end{equation}
which allows to extract an invariant scalar product between Dirac spinors, which we use to orthonormalise the modes.

Having fixed the mode decomposition of the field, we can use it to compute the Wightman function of the Dirac spinors over the Bunch-Davies vaccuum. Consider then
\begin{equation}
\begin{aligned}
	W_{1/2,\frac{d}{2}+i m} &= \bra{0}\psi(x,\eta_1) \overline{\psi}(y,\eta_2)\ket{0} \\
	&=\int \frac{d^{d}p}{(2\pi)^{d}}  e^{ip\cdot (x-y)}\sum_{r}u_{r,p}(\eta_1)\overline{u}_{r,p}(\eta_2) \, .
\end{aligned}
\end{equation}
Replacing the explicit expressions for the polarisations, and using the orthogonality relations for the eigenspinors $\xi$, we find the rather complicated expression 
\begin{equation}
\begin{aligned}
	W_{1/2,\frac{d}{2}+i m} &=\frac{1}{2}(\eta_1\eta_2)^{\frac{d+1}{2}}\int \frac{d^{d}p}{(2\pi)^{d}}  e^{ip\cdot(x-y)}\biggm(\\
	&-i\gamma_0 p\frac{\overline{h}_{im+\frac{1}{2}}(\eta_1 p)h_{im-\frac{1}{2}}(\eta_2p)+\overline{h}_{im-\frac{1}{2}}(\eta_1 p)h_{im+\frac{1}{2}}(\eta_2p)}{2}\\
	&-ip\cdot \gamma \frac{\overline{h}_{im+\frac{1}{2}}(\eta_1 p)h_{im+\frac{1}{2}}(\eta_2p)+\overline{h}_{im-\frac{1}{2}}(\eta_1 p)h_{im-\frac{1}{2}}(\eta_2p)}{2} \\
	&+p\frac{\overline{h}_{im+\frac{1}{2}}(\eta_1 p)h_{im-\frac{1}{2}}(\eta_2p)-\overline{h}_{im-\frac{1}{2}}(\eta_1 p)h_{im+\frac{1}{2}}(\eta_2p)}{2}\\
	&-\gamma_0 p\cdot \gamma \frac{\overline{h}_{im+\frac{1}{2}}(\eta_1 p)h_{im+\frac{1}{2}}(\eta_2p)-\overline{h}_{im-\frac{1}{2}}(\eta_1 p)h_{im-\frac{1}{2}}(\eta_2p)}{2}  \biggm) \, ,
\end{aligned}
\end{equation}
and we will refer to the contributions associated to each line as $I^{(1/2/3/4)}$. These are not as daunting as it may seem, as their evaluation follows from the one of the scalar $I_\nu$, \eqref{eq:fund}. We can extrapolate these integrals using differential relations. Note, from our knowledge of the embedding space, that the results must conspire into the shape 
\begin{align}
	W_{1/2,\frac{d}{2}+im} = \frac{\gamma_a (x-y)^a+\gamma_0\left(\eta_1-\eta_2 \right)}{\sqrt{\eta_1 \eta_2}}g_+(u)+\frac{i\gamma_0\gamma_a (x-y)^a+i\left(\eta_1+\eta_2 \right)}{\sqrt{\eta_1 \eta_2}}g_-(u) \, ,
\end{align}
with $x^{0}=\eta_1$, and $\gamma_0 x^{0}= \gamma_0 \eta_1$. The terms with factors of $ p\cdot \gamma$ are already of the right form to be evaluated using $I_{im \pm \frac{1}{2}}$. One must simply trade $p\cdot \gamma  \equiv -i\gamma \cdot \pdv{}{(x-y)}$, which further simplifies when acting on a function of $u$. This gives

\begin{equation}
\begin{aligned}
	I^{(2)} &= \frac{\gamma_a(x-y)^{a}}{\sqrt{\eta_1\eta_2}}\frac{\partial_u}{8} \left( I_{im-\frac{1}{2}}(u)+I_{im+\frac{1}{2}}(u) \right) \, , \\
	I^{(4)} &= \frac{i\gamma_0\gamma_a(x-y)^{a}}{\sqrt{\eta_1\eta_2}}\frac{\partial_u}{8} \left( I_{im-\frac{1}{2}}(u)-I_{im+\frac{1}{2}}(u) \right) \, .
\end{aligned}
\end{equation}

By itself, these are enough to extrapolate the functions $g_{\pm}$. We can however also evaluate the remaining integrals. The trick is to use the raising operators \eqref{eq:raise}, noting
\begin{align}
	L_{\pm,\nu}^{(\eta)}\left[(-\eta)^{\frac{d}{2}}h_{\nu}(k\eta)\right]&=\pm i \eta k (-\eta)^{\frac{d}{2}} h_{\nu\pm 1}(k\eta) \, ,
\end{align}
and likewise for $\overline{h}$, we can write 
\begin{equation}
	\begin{aligned}
	I^{(1)}=\frac{\gamma_0}{4}\sqrt{\eta_1\eta_2}\left(\frac{1}{\eta_1}L_{+,im-\frac{1}{2}}^{(\eta_1)}-\frac{1}{\eta_2}L_{+,im-\frac{1}{2}}^{(\eta_2)}\right)\left[I_{im-\frac{1}{2}}(u)\right] \, , \\
	I^{(3)}=i\frac{1}{4}\sqrt{\eta_1\eta_2}\left(\frac{1}{\eta_1}L_{+,im-\frac{1}{2}}^{(\eta_1)}+\frac{1}{\eta_2}L_{+,im-\frac{1}{2}}^{(\eta_2)}\right)\left[I_{im-\frac{1}{2}}(u)\right] \, .
\end{aligned}
\end{equation}
It is rather non-trivial that these different expressions combine back to form the precise structure needed, forming a strong check of our computation.

These derivatives give linear combinations of hypergeometric functions with shifted arguments. These can be further simplified by using a variety of hypergeometric identities, or more prosaically by taking the Taylor series, simplifying each terms at a given order, and resumming back. The endpoint of this discussion is that one lands back on the result \eqref{eq:W12} specialised to $\Delta=\frac{d}{2}+im$, 
\begin{equation}
\begin{aligned}
	g_+(z)&= \kappa \, _2F_1\left(\frac{d}{2}+im+\frac{1}{2},\frac{d}{2}-im+\frac{1}{2};\frac{d+1}{2};u\right) \, , \\
	g_{-}(z)&= -\frac{2 i m}{d+1} \kappa \, _2F_1\left(\frac{d}{2}+im+\frac{1}{2},\frac{d}{2}-im+\frac{1}{2};\frac{d+1}{2};u\right) \, , \\
	\kappa &=\frac{\Gamma(\frac{d+1}{2}+im)\Gamma(\frac{d+1}{2}-im)}{2(4\pi)^{\frac{d+1}{2}}\Gamma(\frac{d+1}{2})} \, ,
\end{aligned}
\end{equation}
in perfect agreement. The reader should note that our derivations are completely independent on the value of $d$, and that further requiring reality conditions such as for Majorana particles does not alter the result in any meaningful way. The rederivation of the structures from a mode resummation is an interesting output of this exercise.

Guided by these results, we can consider the harder problem of a generic spinor 2-point function from purely group-theoretical arguments.

\subsection{Deriving the Spectral Decomposition}\label{sec:kl}

Having studied in details the structure of free fields, we know enough to discuss general insertions in a non-perturbative fashion. Our derivation follows the strategy of \cite{Hogervorst:2021uvp}. We consider bulk fields $\psi$ which transforms as a Dirac spinor representation of $SO(1,d+1)$, $\psi(x^\mu)$, with the $SO(1,d)$ subgroup made manifest.  Our goal is to fix the Wightman function $\bra{\Omega}\psi(\eta_1,x)\overline{\psi}(\eta_2,y)\ket{\Omega}$ of the fields $\psi$, over the interacting vacuum $\ket{\Omega}$, using only the symmetry properties of the fields and states. The fields are taken to transform under the isometry group as
\begin{equation}\label{eq:Kf}
\begin{aligned}
	[D,\psi(x^{\mu})] &= x^{\mu}\partial_\mu \psi \, , & [K_a,\psi(x^{\mu})] &= (2x_a x^\mu\partial_\mu - x^\mu x_\mu \partial_a +2\Sigma_{a\mu}x^{\mu})\psi \, , \\
	[P_a,\psi(x^{\mu})] &= \partial_a \psi \, , & 
	[M_{ab},\psi(x^{\mu})] &= -(x_a\partial_b-x_b\partial_a+ \Sigma_{ab})\psi \, .
\end{aligned}
\end{equation}
The Casimir of the conformal group, the operator $\mathcal{C}=D^2-\frac{1}{2}(P\cdot K + K\cdot P)-\frac{1}{2}M_{ab}M^{ab}$, acts on it as 
\begin{align}
	C_{1/2,\Delta}&=\Delta(\Delta-d)+\frac{d(d-1)}{8} \, .
\end{align}

The isometry group also fixes the structure of the Hilbert space. We use this to work in a basis of on-shell states. We pick the states to realise manifestly the $SO(1,1)\times SO(d)$ subgroup. They will be labelled  $\ket{\Delta,k}^{r}$, with $r$ an index of an irreducible representation of $SO(d)$ and $k^{a}$ a momentum label. Without loss of generality, we can pick states which diagonalise the action of the generators $P_{a}$, with eigenvalues $-i k_a$. The action of all the other generators on the states follow directly through dimensional analysis and the commutation relations \cite{Hogervorst:2021uvp}, 
\begin{equation}
\begin{aligned}
	D\ket{\Delta,k}&=-\left(k\cdot \pdv{}{k}+\frac{d}{2}\right)\ket{\Delta,k} \, , \\
	P_a\ket{\Delta,k}&=-ik_a\ket{\Delta,k} \, , \\
	M_{ab}\ket{\Delta,k}&=\left(k_b\pdv{}{k^{a}}-k_a\pdv{}{k^{b}}-\Sigma_{ab}\right)\ket{\Delta,k} \, ,
\end{aligned}
\end{equation}
and finally, 
\begin{equation}\label{eq:Ks}
\begin{aligned}
	K_a\ket{\Delta,k}=&-i\left(  k_a \pdv{}{k}\cdot\pdv{}{k}-2k\cdot\pdv{}{k}\pdv{}{k^a}-d \pdv{}{k^a}+\left(\Delta-\frac{d}{2}\right)^2\frac{k_a}{k^2}\right)\ket{\Delta,k} \\
	&+2i\Sigma_{ab}\left( \pdv{}{k_b}\pm \left(\Delta-\frac{d}{2}\right)\frac{k^b}{k^2}\right) \ket{\Delta,k} \, .
\end{aligned}
\end{equation}

The action of the operator $K_a$ on the state is not fully fixed by the algebra. There is an ambiguity, associated to the shadow transform automorphism, as indicated by the $\pm$ sign. We will now pick the definite $+$ sign, thereby fixing what we mean by the label $\Delta$ and settling on a definite realisation of the group. This peculiarity will turn out to be important later on.

Going back to the object of interest, we now insert a complete set of state in the middle of the correlator. By Wigner-Eckart, only states sitting in a spin-$1/2$ representation can have non-zero overlap with the state created by $\bra{\Omega}\psi$. Our task is to fix by symmetry the wavefunction $\Psi^{r}$ of the field over the state 
\begin{align}
	\bra{\Omega}\psi(\eta,x)\ket{\Delta,k}^{r}=\Psi^{r}(\eta,x,k) \, .
\end{align}
The result for the conjugate wavefunction follows directly from this one and does not carry new information. We proceed with the commutation relations of $P_a$ with the fields to get rid of all position dependance through a Fourier-transform term
\begin{align}
	\Psi^{r}(\eta,x,p)=e^{ik\cdot x}\Psi^{r}(\eta,k)
\end{align} 

The next step is rotational symmetry. Picture using an induced representation for the ket $\ket{\Delta,k}$ to bring ourselves to a given standard momenta $\overline{k}_a=\delta_{a,d} k$. In this frame, we can use rotational symmetry to fix spinning contributions up to a few functions of $\eta$ and $\abs{k}=k$. Once this is done, we can rotate back to obtain an expression valid in any frame. 

From the oscillator construction of the spinor representations, see for example \cite{Polchinski:1998rr}, it is clear that in a Dirac spinor $\psi$ of $Spin(1,d)$  contains two independent pieces transforming as Dirac spinors under $Spin(d)$. Without loss of generality, these can be taken through definite $\gamma_0$ eigenvalue sectors, as they do not mix under spatial rotations. We decompose $\Psi^{r}$ using a basis of $+i, \xi^{r}$, and $-i, \chi^{r}$ eigenvectors of $\gamma_0$. This basis is left unchanged as one goes back from the standard momenta to the generic one. Note furthermore that in the standard frame, we can trade of $\chi \rightarrow \frac{k\cdot \gamma}{k}\xi$, since this is just a basis redefinition. This expression is already covariant, hence is valid in all frames.  All in all, we aruged that rotational symmetry allows us to write
\begin{align}
	\Psi^{r}(\eta,k)= \left(\sqrt{k}f_1(\eta,k)+\frac{ik\cdot \gamma}{\sqrt{k}}f_2(\eta,k) \right)\xi^{r} \, ,
\end{align}
where we already wrote the ansatz in a suggestive way, but without loss of generality.

Our remaining task is to fix the functions $f_1$ and $f_2$. We have not yet made any further assumption apart from translational and rotational symmetry. We will now use more group-theoretic properties of the state. By definition, it must be an eigenvector of the Casimir operator, as is the bulk field. This also means that the wave-function is a solution of the Casimir acting on $\psi$. Knowing the transformation law of $\psi$, we can rewrite the eigenvalue equation for $\ket{\Delta,k}$ as a differential equation for $f_1$ and $f_2$. The output is a set of 2nd order coupled differential equation.  The equations can be decoupled when written in terms of the variables 
\begin{align}
	g_1 &= f_1 + f_2 \, , & g_2 &= f_1 - f_2 \, ,
\end{align}
where they take the form 
\begin{equation}
\begin{aligned}
	\left(\eta^2 \partial_\eta^2 - (d-1)\eta \partial_\eta+ (k^2 \eta^2 + \frac{(d+1)^2}{4}+\left(\Delta(d-\Delta)+k\eta(k\eta-i)\right) \right) g_1(\eta,k) &=0  \, , \\
	\left(\eta^2 \partial_\eta^2 - (d-1)\eta \partial_\eta+ (k^2 \eta^2 + \frac{(d+1)^2}{4}+\left(\Delta(d-\Delta)+k\eta(k\eta+i)\right) \right) g_2(\eta,k) &=0 \, . 
\end{aligned}
\end{equation}

This system is similar to \eqref{eq:dirac1} we had found when canonically quantising the fermion. It can be solved using specific linear combinations of the type of Hankel functions previously used. A peculiar result thus far is that these equations span a four-dimensional parameter space, not two. The solutions for $\Psi$ which are equivalent to the modes $u$ and $v$ previously found, as well as the equivalent modes with $m \rightarrow -m$. All in all, we have so far showed that translations, rotations and the Casimir equation have fixed 
\begin{align}
	\Psi^{r}(\eta,k,x)=e^{ik\cdot x}\left(c(i\nu)u^{+\nu}_{r,k}(\eta)+\tilde{c}(i\nu)u^{-\nu}_{r,k}(\eta)+c^{\sharp}(i\nu)v^{+\nu}_{r,-k}(\eta)+\tilde{c}^{\sharp}(i\nu)v^{+\nu}_{r,-k}(\eta) \right)\, .
\end{align}

The situation is now rather strange. The negative energy modes are not an issue, we can rule them out by the Wightman axiom positive energy condition, like in the scalar case \cite{Hogervorst:2021uvp}. But the doubling of positive energy mode is problematic. From the perspective of massive Dirac fermions, it looks like the bulk field interpolates for twice too many degrees of freedom on-shell. This counting clashes obviously with what one expects from a flat-space limit as well. This is a strong indication that this second set of mode should simply not be there as well. 

To understand why, let us take a step back. These solutions are compatible with the Casimir equation, as well as rotation and translations. However, as is clear from their action on states, all of these actions are invariant under the shadow transform $m\leftrightarrow -m$. It is then expected that this far, both solutions should be allowed. 

However, there is a supplementary, first order, constraint to specify that the state $\ket{\Delta,k}^{r}$ transforms as a state with definite weight $\Delta=\frac{d}{2}+im$, and not its shadow. This is of course precisely the issue that arises from the realisation of $K_a$ on states.  Hence the cure is to impose on our wave-function that it realises special conformal transformations correctly, through the obvious identity 
\begin{align}
	\bra{\Omega}\psi(x,\eta)K_{a}\ket{\frac{d}{2}+im,k}+\bra{\Omega}\big[ K_a,\psi(x,\eta) \big] \ket{\frac{d}{2}+im,k}&=0 \, .
	\end{align}

Using the relations \eqref{eq:Kf} and \eqref{eq:Ks}, we can impose this condition onto a generic polarisation $e^{ik\cdot x}u_{r,k}^{\nu}(\eta)$. The correct realisation of the special-conformal transformations then requires $m=\nu$, thereby ruling out the second set of modes. 

\begin{figure}[h]
	\centering
	\includegraphics[width=0.8\linewidth]{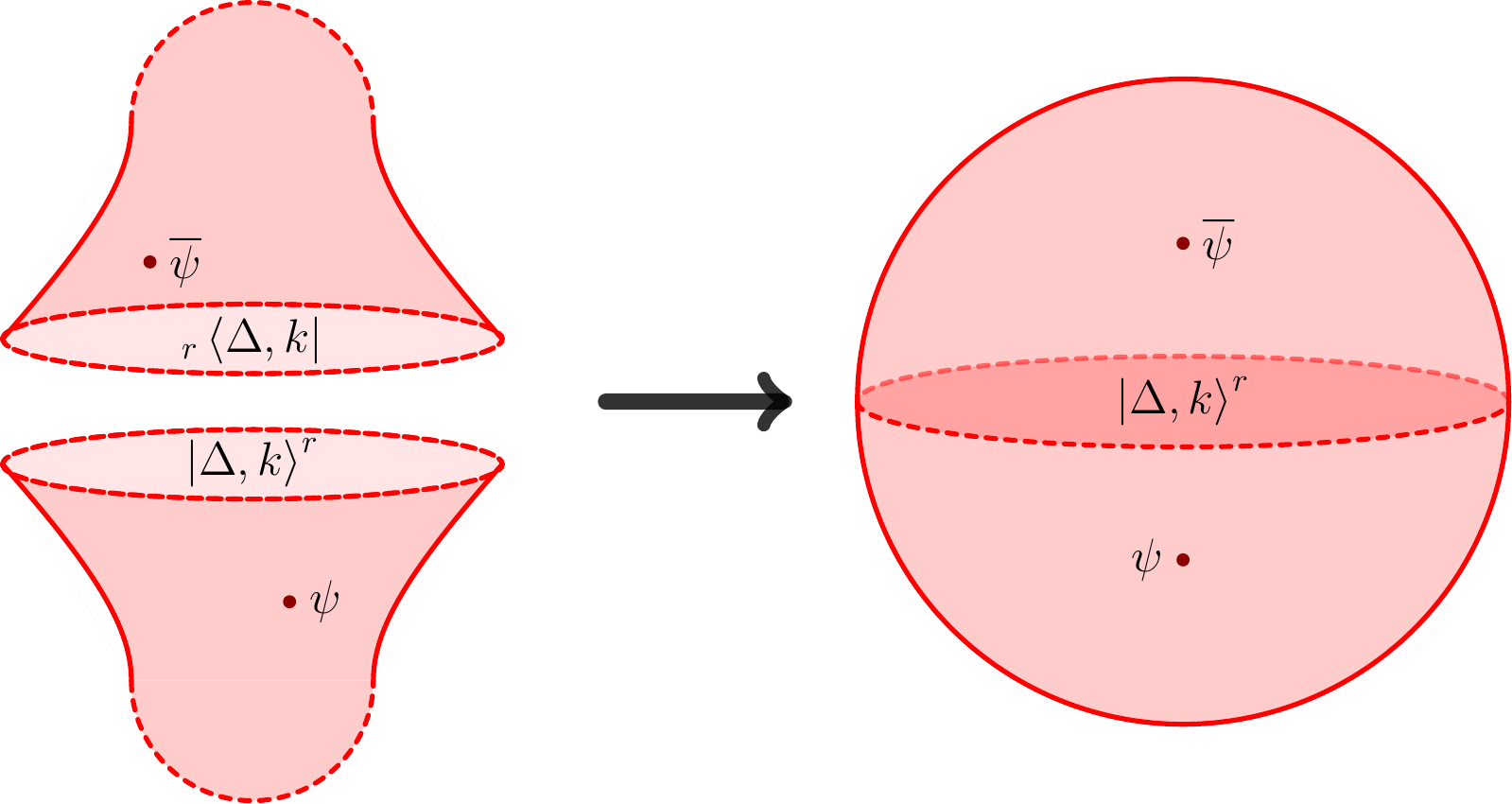}
	\caption{Once the functional form of the correlator in a specific channel is fixed, we can prove its positivity by considering it as two glued copies of de Sitter, whose far-past are regulated by the Bunch-Davies condition to allow for the analytical continuation to the sphere \cite{Schlingemann:1998cw,Schlingemann:1999mk,Higuchi:2010xt}. Glueing the spaces together, we can arrange the points in a symmetric, manifestly positive, configuration, proving the positivity of the coefficient in the expansion.}
	\label{fig:sphere}
\end{figure}

We have shown that the de Sitter isometries require the wave-function of a generic Dirac spinor with a spin-$1/2$ state to be proportional to that of a free massive Dirac spinor mode,
\begin{align}
	\bra{\Omega}\psi(\eta,x)\ket{\frac{d}{2}+i\nu,k}^{r}=c(i\nu)e^{ik\cdot x}u_{r,k}^{\nu}(\eta) \, .
\end{align}
This result is equivalent to the statement that on-shell states satisfy the Dirac equation. A further consequence of which is that these states must all lie on the principal series, with $m\in \mathbb{R}$. This is expected from a group-theoretic perspective, as spin-$1/2$ representation of the de Sitter group sits in the principal series \cite{Thieleker1973,Thieleker1974,Dobrev:1977qv,Joung2006,Joung2007}, but offers a physical derivation of this fact. The result for the conjugate wavefunction follows directly. Bringing these pieces together, we find that the 2-point function admits a decomposition 

\begin{equation}
\begin{aligned}
	\bra{\Omega}\psi(\eta_1,x)\overline{\psi}(\eta_2,y)\ket{\Omega}&= \int_{-\infty}^{\infty} d\nu \rho(i\nu)\int \frac{d^{d}k}{(2\pi)^{d}}  \sum_{r} u^{\nu}_{r,k}(\eta_1)\overline{u}^{\nu}_{r,k}(\eta_2) \\
	&=\int_{-\infty}^{\infty} d\nu \rho(i\nu) W_{1/2,\frac{d}{2}+i\nu}(u) \, .
\end{aligned}
\end{equation}
Where we used back the result previously derived from the mode resummation. The spectral parameter given by 
\begin{align}
	\rho(i\nu)= \frac{\abs{c(i\nu)}^2}{\mathcal{N}}\geq 0 \, ,
\end{align}
is manifestly positive.

An independent argument for the positivity of $\rho$ is obtained through the analytical continuation to the sphere, as illustrated in fig.\ref{fig:sphere}. This concludes our proof.

\section{Discussion}

The stated goal of this work was to tackle cosmological correlators involving fermions. We have reviewed the current understanding for bosonic analytical computations, which revolves around a Wick contraction to Euclidean AdS space. Though they involve some technicalities, we have shown that the same framework accommodates fermions, expanding the previously known transcription rules. This allowed us to import the knowledge developed for Witten diagrams to the evaluation of  perturbative cosmological correlators, as we showed in an example with fermions exchange. Our exposition aimed to show that fermions are no more exotic than bosons in that regard, as the existing strategies for them can be adapted successfully. In the same endeavour, we used the techniques used to study unitarity of some bosonic observables, to prove the analogous statement, at the level of the Källen-Lehman decomposition, for spinors. These results are important as they enrich the observables under analytical control, through which we better our understanding of phenomenologically relevant spacetimes with standard matter content. 

There are many ways through which one can push forward in the future. Much of the literature in de Sitter, such as \cite{anninos2012,anninos2019Holo,David2019,Anninos:2020hfj,Harris2021,Mirbabayi:2022gnl,Muhlmann:2021aa,Anninos2021}, has focused on the properties of the cosmological horizon, often through the angle of the euclidean sphere. We believe an embedding picture can offer insights on this front as well. The associated study of quasinormal modes \cite{Lopez-Ortega:2004vrh,Du:2004jt,Lopez-Ortega:2006aal,Lopez-Ortega:2006tjo,Lopez-Ortega:2012xvr,Jafferis:2013qia,Sun:2020sgn} has already touched on spinors and the embedding picture, but one can make more general statements still. For example, the discussion of \cite{Mirbabayi:2022gnl} can clearly be applied to spinors as well, relating the spectral decomposition we proved to the quasinormal spectrum. A more involved direction would be to study whether the current Euclidean perspective for late-time correlators can be exploited to study the scattering problem in de Sitter, perhaps along the lines of \cite{Albrychiewicz:2020ruh}. Another interesting direction is to make contact with the literature on unitarity and on-shell amplitudes \cite{Meltzer:2019nbs,Goodhew:2020hob,Albayrak:2020fyp,Melville:2021lst,Jazayeri2021,Bonifacio2021,Benincasa:2022gtd,Armstrong:2020woi,Gomez:2021qfd,Armstrong:2022mfr,Armstrong:2022vgl} or the studies of the wavefunctional perspective \cite{Anninos2015,Arkani-Hamed:2017fdk,Cespedes:2020xqq,Salcedo:2022aal}. We believe that on these topics as well, spinors would offer interesting extensions and help better appreciate the structures under study.

\acknowledgments

The author would like to thank  A.~Chalabi, N.~Drukker, E.~Harris, P.~Kravchuk, B.~Pethybridge, and J.~Stedman  for discussions. The author would also like to thank C.~Herzog, D.~Anninos for their comments on the drafts and encouragements. Part of this work was motivated and initiated during the school on ``Black Holes, Cosmology and Holography'' at the STAG Research Centre in Southampton, and during the Corfu Summer Institute ``Workshop on Quantum Features in a de Sitter Universe''. The author is grateful to the organisers of these events. This work was supported by the U.K.\ Science \& Technology Facilities Council Grant ST/P000258/1.

\appendix

\section{Spinorial Harmonic Functions}\label{app:harmonic}

Harmonic functions are useful tools for the evaluation of Witten diagrams. The relevant results for bosonic propagators can be found in \cite{Costa:2014kfa}, while the spinorial variant was introduced in \cite{Nishida:2018opl}. However, the latter is not very detailed, and uses conventions which are very different from ours. It is useful then to study these object in the setup most suited to our concerns.

The goal of this section is to detail the decomposition of the propagator in harmonics. We start by proving the orthonormality property of the spinorial harmonics,
\begin{align}
	\int_{AdS} dx \; \Omega_{1/2,\nu}(x_1;x)\Omega_{1/2,\nu'}(x;x_2)= \delta(\nu-\nu')\Omega_{1/2,\nu}(x_1;x_2) \, .
\end{align}
We use the explicit real-space expression
\begin{align}
	\Omega_{1/2,\nu}(x_1;x_2)&= N_{h+i\nu}N_{h-i\nu}\frac{i}{2\pi} \int d^{d}y  \, \frac{\gamma\cdot(x_1-y)2(1\pm\gamma_0)\gamma\cdot(y-x_2)}{\sqrt{z_1 z_2}\left(\frac{(x_1-y)^2}{z_1}\right)^{h+i\nu+\frac{1}{2}}\left(\frac{(x_2-y)^2}{z_2}\right)^{h-i\nu+\frac{1}{2}} }  \, ,
\end{align}
which is invariant under the replacement $\nu,\gamma_0 \rightarrow -\nu,-\gamma_0$, and we abbreviated $h=\frac{d}{2}$. The integral we want to compute takes the unwieldy form 
\begin{equation}
\begin{aligned}
	&\frac{i}{2\pi}\frac{N_{h+i\nu}N_{h-i\nu}}{\sqrt{z_1 z_2}} \int d^{d}y_1  \,  \frac{i}{2\pi}4 N_{h+i\nu'}N_{h-i\nu'}\int \frac{d^d x dz}{z^{d+1}} \int  d^{d}y_2  \times \\ 
	& \frac{\gamma\cdot(x_1-y_1)(1+\gamma_0)\gamma\cdot(y_2-x)\gamma\cdot(x-y_2)(1-\gamma_0)\gamma\cdot(y_2-x_2)}{z\left(\frac{(x_1-y_2)^2}{z_1}\right)^{h+i\nu+\frac{1}{2}}\left(\frac{(y_2-x)^2}{z}\right)^{h-i\nu+\frac{1}{2}}\left(\frac{(x-y_2)^2}{z}\right)^{h-i\nu+\frac{1}{2}}\left(\frac{(y_2-x_2)^2}{z_2}\right)^{h+i\nu+\frac{1}{2}}} \, .
\end{aligned}
\end{equation}

We will concentrate first on the middle integral over $AdS$, which sits between two boundary integrals. After simplification, this integral can be evaluated similarly to the ones encountered for bosonic harmonic functions, 
\begin{equation}
\begin{aligned}
	&\int \frac{d^{d}x \, dz}{z^{d+1}}\frac{(1+\gamma_0)(\gamma\cdot(y_1-x)\gamma\cdot(x-y_2))(1-\gamma_0)}{z\left(\frac{(x-y_1)^2}{z}\right)^{h+\frac{1}{2}-i\nu}\left(\frac{(x-y_2)^2}{z}\right)^{h+\frac{1}{2}-i\nu'}} \\
	=&2(1+\gamma_0)\gamma\cdot(y_1-y_2) \int \frac{d^{d} x  \, dz}{z^{d+1}}\frac{1}{\left(\frac{(x-y_1)^2}{z}\right)^{h+\frac{1}{2}-i\nu}\left(\frac{(x-y_2)^2}{z}\right)^{h+\frac{1}{2}-i\nu'}} \\
	=& (1+\gamma_0)\gamma\cdot(y_1-y_2)\left( \frac{4\pi\sqrt{\pi}^{d}\Gamma(\frac{1}{2}-i\nu)\delta(\nu-\nu')}{\Gamma(\frac{d+1}{2}-i\nu)\left((y_1-y_2)^2\right)^{h+\frac{1}{2}-i\nu}}+\delta^{d}(y_1-y_2)\ldots \right) \, .
\end{aligned}
\end{equation}
In the last line we used the integral identity derived in the appendix of \cite{Costa:2014kfa}. The second piece in the last expression does not contribute because of the spinorial structure. We are left with an integral over the point $y_2$, which is effectively a shadow transform changing the last power law from $h+\frac{1}{2}+i\nu$ to $h+\frac{1}{2}-i\nu$. This integral is tedious to perform in position space, and so we rewrite the expression back to the embedding space,
\begin{equation}
\begin{aligned}
	& \int d^{d}y_2 \frac{\gamma\cdot(y_1-y_2)\gamma\cdot(y_2-x_2)}{((y_1-y_2)^2)^{h+\frac{1}{2}-i\nu}\left(\frac{(y_2-x_2)^2}{z_2}\right)^{h+\frac{1}{2}+i\nu}} \\
	=& \int [dP_2] \frac{\overline{S}_{1,\partial}[y_1]\slashed{P}_2S[x_2]}{(-2P_1\cdot P_2)^{h+\frac{1}{2}-i\nu}(-2P_2\cdot X)^{h+\frac{1}{2}+i\nu}} \\
	=&\frac{\overline{S}_{1,\partial}[y_1] \Gamma_A X^A S[x_2]\sqrt{\pi}^{d}\Gamma(\frac{d}{2}+1)}{\Gamma(h+\frac{1}{2}+i\nu)\Gamma(h+\frac{1}{2}-i\nu)}\int_{0}^{\infty} d\lambda \frac{\lambda^{\frac{d-1}{2}+i\nu}}{\big(1+\lambda(-2X\cdot P_1) \big)^{\frac{d+2}{2}}} \\
	=&(-i) \frac{\pi ^{d/2} \Gamma \left(i \nu +\frac{1}{2}\right)}{\Gamma(\frac{d+1}{2}+i \nu)}\frac{\overline{S}_{1,\partial}[y_1] S[x_2]}{(-2 X\cdot P_1)^{h-i\nu+\frac{1}{2}}} \, .
\end{aligned}
\end{equation}

 The remaining integral then joins together the bulk-to-boundary propagator, giving back $\Omega_{1/2,\nu}$ and the $\delta(\nu-\nu')$ needed. One must simply be careful to piece together the constants. Once the dust settles we are left with the final result 
 \begin{align}
	\int_{AdS} d^{d+1}x \, \Omega_{1/2,\nu}(x_1; x)\Omega_{1/2,\nu'}(x;x_2)= \delta(\nu-\nu')\Omega_{1/2,\nu}(x_1;x_2) \, .
\end{align}
This computation implies the identity 
\begin{align}\label{}
	\int d\nu \; \Omega_{1/2,\nu}(x;y)= \delta_{H}^{d+1}(x-y) \, .
\end{align}

This completeness relation implies one can write a generic propagators using a superposition of harmonics i.e. the propagator admits a decomposition
\begin{align}
	\Pi^{+}_{h+m}(x;y)=-i\int \frac{d\nu'}{\nu'+i m} \Omega_{1/2,\nu'}(x;y) \, .
\end{align}
The integration kernel is fixed using the equations of motions. Since we know that the difference of the two propagators is proportional to the harmonic function, we can solve for $\Pi^{-}_{h-m}$. These expressions are valid provided $Re[m]>0$, hence to obtain $\Pi^{+}_{h+i\nu}$ we consider $m=\epsilon+i\nu$, $\epsilon>0$, and take the limit to the principal series
\begin{align}
	\Pi^{+}_{h+i\nu}= -i \int \frac{d\nu'}{\nu'-\nu+i\epsilon}\Omega_{1/2,\nu'}(x;y)  \, .
\end{align}
The analogous expression for $\Pi^{-}_{h-i\nu}$ follows through by using the Sokhotski-Plemelj formula. The final result is particularly appealing
\begin{align}
	\Pi^{\pm}_{h\pm i\nu}(x;y)&=-i\int \frac{d\nu'}{\nu'-\nu\pm i\epsilon} \Omega_{1/2,\nu'}(x;y) \, .
\end{align}
From this, we gather that to evaluate a general diagram involving any type of fermionic propagator, we simply have to replace each propagator by an harmonic function with arbitrary weight, and integrate over it with the appropriate kernel.

\bibliographystyle{jhep}
\bibliography{paper.bib}

\end{document}